\documentclass[onecolumn,journal]{IEEEtran}
\usepackage[utf8]{inputenc}
\usepackage[T1]{fontenc}
\usepackage{natbib}
\setcitestyle{authoryear,maxnames=1}    
\usepackage{amsmath,amssymb,amsfonts,mathtools,bm}
\usepackage{url}
\usepackage{eurosym}
\usepackage{tabularx}
\usepackage{longtable}
\usepackage{float}
\usepackage[usenames,dvipsnames]{xcolor}
\usepackage{graphicx}
\usepackage{algorithm}
\usepackage{algpseudocode}
\usepackage{array}
\usepackage[caption=false,font=normalsize,labelfont=sf,textfont=sf]{subfig}
\usepackage{textcomp}
\usepackage{stfloats}
\usepackage{verbatim}
\usepackage{tikz}
\usepackage{siunitx}
\usepackage{arydshln}
\usepackage{balance}
\usepackage{booktabs}
\usepackage{bbold}
\usepackage{multirow}
\usepackage{marginnote}
\usepackage{color, soul}
\usepackage{multicol}
\usepackage[makeroom]{cancel}
\usepackage{cases}

\newtheorem{assumption}{Assumption}
\newtheorem{lemma}{Lemma}
\newtheorem{theorem}{Theorem}

\newtheorem{corollary}{Corollary}

\makeatletter
\renewcommand\appendix{
    \par
    \setcounter{section}{0}
    \setcounter{subsection}{0}
    \gdef\thesection{\Alph{section}}
}
\makeatother

\begin{document}
	
\title{Goal-oriented safe active learning for predictive control using Bayesian recurrent neural networks}
	
\author{Laura Boca de Giuli$^1$, Alessio La Bella$^1$, Manish Prajapat$^2$, Johannes K\"ohler$^3$, Anna Scampicchio$^4$, Riccardo Scattolini$^1$, and Melanie Zeilinger$^2$
\thanks{$^1$Laura Boca de Giuli, Alessio La Bella, and Riccardo Scattolini are with the Department of Electronics, Information, and Bioengineering, Politecnico di Milano, Milan, Italy (e-mails: \textsl{laura.bocadegiuli@polimi.it}, \textsl{alessio.labella@polimi.it}, \textsl{riccardo.scattolini@polimi.it}). \\
	$^2$Manish Prajapat and Melanie Zeilinger are with the Institute for Dynamic Systems and Control, ETH Z\"urich, Z\"urich, Switzerland (e-mails: \textsl{manishp@ethz.ch}, \textsl{mzeilinger@ethz.ch}). \\
	$^3$Johannes K\"ohler is with the Department of Mechanical Engineering, Imperial College London, London, UK (e-mail: \textsl{j.kohler@imperial.ac.uk}). \\
	$^4$Anna Scampicchio is with the Department of Electrical Engineering, Chalmers University of Technology, G\"oteborg, Sweden (e-mail: \textsl{anna.scampicchio@chalmers.se}).}}
	
\maketitle

\begin{abstract}
A key challenge in learning-based model predictive control (MPC) is to collect informative data online for model adaptation while ensuring safety and without penalising control performance. 
In this paper, we propose an online model adaptation scheme embedded within an MPC framework in which the last-layer parameters of a recurrent neural network are recursively updated via Bayesian learning. This is achieved by means of a goal-oriented safe active learning algorithm that alternates between an exploration phase, where the MPC actively explores system dynamics to collect informative data for model adaptation while still pursuing the main control objective, and a goal-reaching phase, where it focuses exclusively on the main control objective. The algorithm is complemented with theoretical guarantees of \textit{(i)} recursive feasibility, \textit{(ii)} safety, \textit{(iii)} termination of exploration in finite time, and \textit{(iv)} close-to-optimal performance. Simulation results on a benchmark energy system demonstrate that the proposed framework achieves economic performance comparable to that of an MPC with full system knowledge, while progressively improving model accuracy and respecting operational safety constraints with high probability.
\end{abstract}


\section{Introduction}
\label{sec:intro}
Model predictive control (MPC) has become one of the most widely adopted strategies for the control of complex dynamical systems subject to constraints~\citep{garcia1989model}. As MPC is typically designed under the certainty equivalence principle~\citep{james1994certainty}, its performance strongly depends on the accuracy of the underlying prediction model. In recent years, data-based models such as neural networks (NNs)~\citep{anderson1995introduction} and recurrent neural networks (RNNs)~\citep{medsker2001recurrent} have gained increasing attention within the MPC community, owing to the vast availability of data and their relatively low development effort compared to physics-based models~\citep{morato2024data}. However, the predictive capability of black-box models is highly dependent on the quality and coverage of the training dataset, which often does not span the full range of operating conditions encountered during real-world operation~\citep{gawlikowski2023survey}. As a result, models trained offline may lead to significant degradation in control performance when deployed during operation. It is therefore desirable to continuously adapt these models using online data, as the offline training dataset may be insufficient~\citep{harrison2018meta,mckinnon2019learn}. At the same time, data collected during closed-loop operation may not be sufficiently informative unless the system is actively excited. Actively exciting the system to gather informative data can thus improve model adaptation~\citep{camilleri2022active}, but it should be done while respecting safety constraints for the unknown system and while limiting exploration to what is necessary to achieve the control objectives, since exploring the full system dynamics may be inefficient~\citep{mesbah2018stochastic}. In view of this, the objective of this work is to develop an MPC-based algorithm that enables the progressive refinement of RNN parameters over time, while ensuring optimal and safe system operation.

\subsection{Related work}
Online adaptation of data-based models requires the explicit representation and update of an uncertainty component. Examples of some approaches enabling this include: augmenting the nominal model with an auxiliary one~\citep{de2024lifelong}; combining multiple deterministic models via ensemble techniques~\citep{zhang2012ensemble}; exploiting adaptive set-membership estimation~\citep{lorenzen2019robust}, Gaussian processes (GPs)~\citep{williams2006gaussian,scampicchio2025gaussian}, or Bayesian neural networks (BNNs)~\citep{jospin2022hands,fortunato2017bayesian,tran2019bayesian}. GPs have been widely studied due to their ability to directly quantify model uncertainty, which is crucial to ensure safety~\citep{scampicchio2025gaussian}. However, performing inference with full GPs is computationally challenging, especially in an online learning scheme, and approximation techniques aimed at enhancing scalability lack the uncertainty bounds needed for safety certification~\citep{liu2020gaussian}. BNNs represent a valid alternative, as they combine uncertainty quantification with the effectiveness of neural networks by assigning probability distributions to all model parameters and updating them online based on system measurements~\citep{gawlikowski2023survey}. Nevertheless, both training and inference in fully Bayesian models are typically computationally intractable~\citep{fiedler2022model}. A promising approximation that mitigates these computational challenges is the Bayesian last-layer (BLL) approach, in which only the parameters of the output layer of the network are treated as uncertain variables. As a result, the computational complexity scales with the number of last-layer parameters rather than with the amount of training data as in GPs~\citep{fiedler2023improved, tran2019bayesian}. 

While several MPC frameworks based on BNNs and BLLs have been proposed~\citep{fiedler2022model,pohlodek2023stochastic,johnson2024robust}, they update their models using available data without assessing their relevance. To effectively reduce model uncertainty, it is instead essential to actively operate the system in a way that promotes the collection of informative data. This has motivated the development of methods of active learning, in which the system is optimally controlled while simultaneously exploring its dynamics~\citep{soloperto2020augmenting,heirung2018model,li2024dual,mesbah2018stochastic,cohn1993neural,jain2018learning}. Most approaches achieve this by maximising an information gain term within the MPC cost function~\citep{heirung2015mpc}. However, many active learning methods do not provide safety guarantees~\citep{thangavel2018dual,iannelli2020structured}, whereas integrating active learning with safety mechanisms is crucial to ensure that operational constraints are satisfied while uncertain system dynamics are being explored, as evidenced in~\citep{camilleri2022active,ewen2023not,schneider1996exploiting}. These approaches can ensure safe operation, but they typically lack theoretical guarantees that the learning process terminates in finite time and leads to close-to-optimal performance. In fact, learning should be limited to what is necessary to achieve the control objective, as attempting to identify the full system dynamics can be inefficient and may interfere with the main control objective. A safe active learning framework with finite exploration is discussed in~\citep{prajapat2025safe}, employing GPs to learn unknown system constraints up to a prescribed confidence level. Considering unknown system dynamics, \citep{prajapat2025safe2} proposes a GP-based approach with finite and close-to-optimal exploration; however, it requires computationally expensive estimation of the reachable set for the uncertain state trajectories to ensure safety. BLL-based approaches, in contrast, do not introduce additional computational burden due to their efficient online updates and do not require intensive estimation of reachable state-trajectories set, as the uncertainty is confined to the output layer. In this context, a safe active learning framework based on the online adaptation of the BLL of a feedforward neural network is proposed in~\citep{lew2022safe}; however, it does not provide guarantees of finite exploration or performance optimality.

\subsection{Main contribution}
In light of the above discussion, this paper builds on the previous results in \citep{prajapat2025safe2, lew2022safe} to establish an active learning control framework based on a BLL-RNN that is computationally efficient, safe, and goal-oriented. Specifically, we make the following contributions:
\begin{enumerate}
    \item \textit{Recursive online update of BLL parameters of an RNN with safety guarantees}: We quantify and update the uncertainty of an RNN model online using a Bayesian last-layer approach. The output-layer parameters are recursively updated using informative data collected via an MPC regulator, which learns system output up to a predefined confidence level while controlling the system to achieve the main objective. Safety is ensured by enforcing online-adapted conservative bounds, thereby guaranteeing that all operational constraints are satisfied by the unknown system output with high probability.
    \smallskip
    \item \textit{Goal-oriented safe active learning with finite exploration and close-to-optimal performance}: We devise a goal-oriented safe active learning algorithm in which the controller switches in finite time from exploration to goal-reaching control once the BLL parameters have been learned to the extent required to achieve the control objective. The switching criterion is based on a comparison between the costs obtained by a pessimistic MPC formulation, employing cautious constraints, and an optimistic formulation with confident constraints. We complement the algorithm with theoretical guarantees ensuring, with high probability, that: \textit{(i)} all MPC optimisation problems are recursively feasible; \textit{(ii)} safety constraints are always satisfied by the unknown system; \textit{(iii)} exploration terminates in finite time; and \textit{(iv)} the closed-loop performance upon termination of exploration is close to that of an MPC with full system knowledge.
\end{enumerate}

Compared with existing works \citep{prajapat2025safe2, lew2022safe}, the proposed framework based on BLL-RNNs offers several key advantages. First, restricting uncertainty to the linear output layer of an RNN, which is recursively updated via Bayesian learning, avoids computationally demanding reachable-set calculations and enables continuous output model refinement without the need to store all past data. Second, the proposed finite, goal-oriented exploration algorithm prevents indefinite exploration that could otherwise interfere with control performance, learning only what is necessary to achieve close-to-optimal performance. Consequently, after the exploration phase, the controller focuses entirely on goal-reaching, achieving performance comparable to an MPC with full system knowledge. Finally, all of this is achieved while always respecting operational safety constraints thanks to the cautious constraint formulation within the MPC.

The proposed algorithm is validated on a benchmark district heating system~\citep{krug2021nonlinear}, where the objective is to learn the model outputs online while respecting operational safety constraints and minimising production costs. Simulation results demonstrate that the algorithm progressively refines model predictions, respects all operational safety constraints, terminates exploration in a finite number of steps, and achieves economic performance comparable, in terms of cost savings, to that of an MPC regulator with full system knowledge.

\subsection{Paper outline}
This paper is organised as follows. Section~\ref{sec:problemstat} presents the problem formulation and the objectives of the work. Section~\ref{sec:exploration} introduces the Bayesian last-layer methodology and formulates the MPC problem for safe active learning. Section~\ref{sec:goalor} introduces the proposed goal-oriented safe active learning algorithm with finite exploration and close-to-optimal performance guarantees. Simulation results are presented in Section~\ref{sec:casestudy}. Final considerations are given in Section~\ref{sec:conclusions}.

\subsection{Notation}
Let $\mathbb{N}$ denote the set of natural numbers, $\mathbb{N}_{>0}$ the set of positive integers, $\mathbb{R}$ the set of real numbers, and $\mathbb{R}_{>0}$ the set of positive real numbers. The Hadamard (element-wise) product between the vectors $z$ and $v$ is denoted by $z \circ v$. The ball of radius $\epsilon$ is defined as $\mathcal{B}_{\epsilon} = \{ z \in \mathbb{R}^n : \lVert z \rVert_2 \leq \epsilon \}$. For a symmetric matrix $A$, $\overline{\lambda}(A)$ and $\underline{\lambda}(A)$ denote its maximum and minimum eigenvalues, respectively, $A\succ 0$ and $A\succeq 0$ indicate that $A$ is positive definite or semidefinite, respectively, and $\lVert z \rVert_A = \sqrt{z^{\top} A z}$. The identity matrix of dimension $n$ is denoted as $I_n$. Sets are denoted by calligraphic letters, e.g., $\mathcal{X}$. The Pontryagin set difference for two sets $\mathcal{X}, \mathcal{Y} \subseteq \mathbb{R}^n$ is defined as $\mathcal{X} \ominus \mathcal{Y} \coloneq \{z \in \mathbb{R}^n \mid z+y \in \mathcal{X} \; \forall y \in \mathcal{Y} \}$. The probability of an event $A$ is denoted by $\mathbb{P}(A)$.

\section{Problem statement}
\label{sec:problemstat}
Consider a discrete-time system
\begin{subequations}
\label{eq:sys}
	\begin{numcases}{}
    \begin{aligned}
    \label{eq:x_sys}
		x_{k+1} = \phi(x_k, u_k)
    \end{aligned}\\
    \begin{aligned}
    \label{eq:y_sys}
		\,\,\, y^\star_k \,\:\: = \theta^{\star \top} x_k 
    \end{aligned}
	\end{numcases}
\end{subequations}
where $x_k\in \mathbb{R}^{n_x}$, $u_k \in \mathbb{R}^{n_u}$, and $y_k^\star \in \mathbb{R}^{n_y}$ represent the state, input, and output of the system, respectively, and $k \in \mathbb{N}_{>0}$ is the discrete-time index. The system is subject to the nonlinear dynamics \mbox{$\phi : \mathbb{R}^{n_x+n_u} \rightarrow \mathbb{R}^{n_x}$}~\eqref{eq:x_sys}, while the output is a linear transformation of the state~\eqref{eq:y_sys}, where $\theta^\star \in \mathbb{R}^{n_x \times n_y}$ denotes the output parameter matrix. Although this model structure is representative of a broad class of dynamical systems, we assume here that~\eqref{eq:sys} corresponds to a recurrent neural network (RNN), which is particularly suited for representing dynamical systems~\citep{bonassi2022recurrent}. Therefore, $u$ and $y$ represent physical quantities, whereas $x$ is the hidden state of the RNN. 
The measured output is denoted as 
\begin{align}
\label{eq:measuredoutput}
    \tilde{y}^\star_k = y^\star_k + \eta_k,
\end{align}
where $\eta \in \mathbb{R}^{n_y}$ is the measurement noise, assumed to be conditionally $\sigma$-subGaussian, including the case of
zero-mean noise bounded in $[-\sigma,\sigma]$~\citep[Eq.(1)]{chowdhury2017kernelized}. For clarity of notation, we assume $n_y=1$; for $n_y>1$, all output-related expressions hold element-wise. 

We assume that the state dynamics $\phi$ are known (i.e., given by the pre-trained RNN) and that the state $x$ is accessible, as standard in last-layer adaptation methods for NNs. The output parameters $\theta^{\star}$ are unknown and must be learned from data during operation. Clearly, the exact $\theta^{\star}$ cannot be learned from a finite number samples; therefore, we aim to learn them up to a specified level of confidence. To this end, at each time instant, we update the output model $y_k = \theta^{\top} x_k$, where $\theta$ denotes the estimate of $\theta^{\star}$. 

The problem addressed in this paper is to optimally control system~\eqref{eq:sys} with respect to a desired objective, i.e., minimising the generic time-varying cost $\mathcal{L}_k(y)$ -- such as economic cost or reference tracking -- while simultaneously learning the output model and ensuring that operational safety constraints are satisfied, i.e., imposing that
\begin{align}
    x_k \in \mathcal{X}, \;\;\; u_k \in \mathcal{U}, \;\;\; \forall k \in \mathbb{N}_{>0},
    \end{align}
and, being the real output uncertain, that    
\begin{align}
\label{eq:output_constr}
     \mathbb{P}(y^\star_k \in \mathcal{Y}\coloneq [y_{\mathrm{min}}, y_{\mathrm{max}}], \, \forall k \in \mathbb{N}_{>0})\geq 1-\delta,
\end{align}
where $\delta \in (0,1)$ is typically small so that \eqref{eq:output_constr} holds with high probability.  

Ultimately, this work addresses two main objectives:
\begin{itemize}
    \item Objective 1: learning output model parameters during operation by collecting informative data, while guaranteeing satisfaction of safety constraints for the unknown system output and controlling it according to the main control objective;
    \smallskip
    \item Objective 2: guaranteeing that dynamics exploration terminates in finite time, learning output model parameters to the extent necessary to achieve close-to-optimal control performance.
\end{itemize}

\section{Objective-aware safe active learning via Bayesian last layer}
\label{sec:exploration}
In this section, we address Objective 1. In Section~\ref{subsec:bll}, we introduce the basic notion of Bayesian linear regression used to update the last layer of an RNN model online. In Section~\ref{subsec:objawareexplo}, we present the MPC module responsible for the objective-aware safe active learning strategy, which is designed to explore informative system dynamics while simultaneously respecting safety constraints and pursuing the main control objective, hence the name \textit{objective-aware safe active learning}. Thus, parameters are learned up to a user-defined tolerance $\epsilon \in \mathbb{R}_{>0}$, as the exact parameters cannot be learned from a finite number of samples~\citep{prajapat2025safe}.

\subsection{Bayesian last layer}
\label{subsec:bll}
To update the output parameters of the RNN~\eqref{eq:sys}, we follow the Bayesian last-layer (BLL) approach~\citep{fiedler2023improved,tran2019bayesian}, according to which, thanks to the linearity of the output layer of most RNNs (e.g., LSTM, GRU, NNARX, ESN~\citep{bonassi2022recurrent}), the parameters can be efficiently updated online via Bayesian linear regression. Specifically, consider the output-layer parameters characterised by an initial guess for the mean $\bar{\theta}_{0} \in \mathbb{R}^{n_x}$ and for the covariance matrix $\sigma^2 \Lambda_{0}^{-1} \in \mathbb{R}^{n_x \times n_x}$, with $\Lambda_{0}\succ 0$. For our theoretical analysis, we do not need to assume a known prior distribution over the parameters $\theta \sim \mathcal{N}(\bar{\theta}_{0},{\sigma}^2\Lambda_{0}^{-1})$ or Gaussian noise. 
Instead, similarly to~\citep{lew2022safe}, we consider more general $\sigma$-subGaussian noise and the following (mild) bound on the initial guess. 
\begin{assumption} 
\label{ass:prior}
The prior parameters $\bar{\theta}_{0}$ and $\Lambda_{0}$ are such that $\lVert \theta^{\star} - \bar{\theta}_0 \rVert_{\Lambda_0}^2 \leq C$, with known constant $C \in \mathbb{R}_{>0}$.
\end{assumption}

The posterior mean and covariance of the parameters are updated recursively at each time index $k$ as~\citep{harrison2018meta}
\begin{subequations}
\label{eq:update_param1}
	\begin{align}
    \Lambda_{k}^{-1} & = \Lambda_{k-1}^{-1} - \frac{(\Lambda_{k-1}^{-1} x_k)(\Lambda_{k-1}^{-1} x_k)^{\top}}{1+x_k^{\top} \Lambda_{k-1}^{-1} x_k}, \label{subeq:lambdakinv} \\
    Q_{k} & = x_k \tilde{y}_k^{\star} + Q_{k-1}, \\
    \bar{\theta}_{k} & = \Lambda_{k}^{-1} Q_{k},  \label{subeq:thetak}
	\end{align}
\end{subequations}
with $Q_{0} = \Lambda_{0} \bar{\theta}_{0}$. As a result, the posterior mean and covariance of the output is given by
\begin{subequations}
\label{eq:update_y}
	\begin{align}
    \mu_k(x) & = \bar{\theta}_{k}^{\top} x, \label{subeq:muk} \\
    \Sigma_k^2(x) & = x^{\top} \sigma^2 \Lambda_{k}^{-1} x. \label{subeq:Sigmak2}
	\end{align}
\end{subequations}

With this BLL update rule, the computational complexity scales with the number of states in the output layer, thus enabling a more computationally efficient online update of the RNN model compared with GPs where the covariance computation scales with the number of training samples~\citep{fiedler2022model}.

Overall, to ensure effective parameter learning, it is important to actively explore the system in a way that generates informative data.

\subsection{Objective-aware safe active learning}
\label{subsec:objawareexplo}
In the following, we propose an active learning strategy through a soft constraint which encourages to explore informative and uncertain output operating regions up to a predefined threshold $\epsilon \in \mathbb{R}_{>0}$, while satisfying operational safety constraints and accounting for the main control objective. This formulation will enable to automatically deactivate the exploration constraint once sufficient information has been gathered to achieve close-to-optimal control behaviours, thereby focusing solely on the main control objective (see Section~\ref{sec:goalor}).

First, similarly to~\citep{lew2022safe}, we introduce the following Lemma, whose proof is deferred to Appendix~\ref{appendix:beta}.

\begin{lemma}
\label{lemma:beta_k}
Let Assumption~\ref{ass:prior} hold, $C$ be the constant introduced in Assumption~\ref{ass:prior}, and consider the updates~\eqref{eq:update_param1}-\eqref{eq:update_y} with system~\eqref{eq:sys} and some probability level $\delta \in (0,1)$. Define
\begin{align}
    \label{eq:beta}
    \beta_{k} = \sqrt{2 \log \left( \frac{1}{\delta} \frac{\det(\Lambda_k)^{1/2}}{\det(\Lambda_0)^{1/2}} \right)} + \sqrt{\frac{C}{\sigma^2}},
\end{align}
and the uncertainty $w_k(x) = \beta_{k} \Sigma_k(x)$. Then it holds that
\begin{equation}
    \label{eq:lemma1}
    \mathbb{P}(\lvert \theta^{\star \top} x - \mu_k(x) \rvert \leq w_k(x), \, \forall k \in \mathbb{N}_{>0}) \geq 1-\delta.
\end{equation}
\end{lemma}
This result provides a high-probability error bound on the difference between the real output $y^{\star}$ and the mean $\mu$ of the estimated output, given by the uncertainty $w$.

Moreover, we define the following lower and upper bounds for the output:
\begin{subequations}
\label{eq:bounds}
	\begin{align}
    \operatorname{lb}_k(x) & = \max \{\operatorname{lb}_{k-1}(x), \mu_{k}(x)-w_k(x) \},  \\
    \operatorname{ub}_k(x) & = \min \{\operatorname{ub}_{k-1}(x), \mu_{k}(x)+w_k(x) \},
	\end{align}
\end{subequations}
where $\operatorname{lb}_0(x)=\mu_0 - \beta_{0} \Sigma_{0}(x)$, $\operatorname{ub}_0(x)=\mu_0 + \beta_{0} \Sigma_{0}(x)$. Note that $\operatorname{lb}_k(x) \leq \operatorname{lb}_{k+1}(x)$ and $\operatorname{ub}_k(x) \geq \operatorname{ub}_{k+1}(x)$ by construction. Thanks to this, we can state the following corollary, whose proof follows straightforwardly from that of Lemma~\ref{lemma:beta_k}.
\begin{corollary}
\label{corollary:beta}
        Let conditions in Lemma~\ref{lemma:beta_k} hold, then $\mathbb{P}(\theta^{\star \top} x \in [\operatorname{lb}_k(x), \operatorname{ub}_k(x)], \, \forall k \in \mathbb{N}_{>0})\geq 1-\delta$.
\end{corollary}

Considering the operational safety constraint~\eqref{eq:output_constr}, the lower and upper bounds for the output enable us to define the pessimistic state set as
\begin{align}
\label{eq:Xkp}
    \mathcal{X}_k^{\mathrm{p}} & = \{x \in \mathcal{X} \mid y_{\mathrm{min}} \leq \operatorname{lb}_k(x) \leq \operatorname{ub}_k(x) \leq y_{\mathrm{max}} \}, 
\end{align}
which will be included in the MPC formulation to ensure that the operational safety constraints are satisfied with high probability at all times by the real system output. In fact, if $x \in \mathcal{X}_k^{\mathrm{p}}$, then, by Corollary~\ref{corollary:beta}, it follows that $y^{\star} \in \mathcal{Y} = [y_{\mathrm{min}}, y_{\mathrm{max}}]$ with probability greater than $1-\delta$, thus satisfying \eqref{eq:output_constr}.

For recursive feasibility, the proposed MPC scheme also relies on a terminal set~\citep{rawlings2017model}.

\begin{assumption}
\label{ass:invset}
At each $k \in \mathbb{N}_{>0}$, there exists a terminal set $\mathbb{X}_k$ which satisfies 
\begin{enumerate}
    \item safety: $\mathbb{X}_k \subseteq \mathcal{X}_k^{\mathrm{p}}$;
    \item monotonicity: $\mathbb{X}_k \subseteq \mathbb{X}_{k+1}$;
    \item invariance: there exists a terminal control action $u_f : \mathbb{X}_k \rightarrow \mathcal{U} \ominus \mathcal{B}_{2 \epsilon}$ such that $\mathbb{X}_k$ is a positive invariant set, i.e., $\forall x \in \mathbb{X}_k$, $\phi(x, u_f(x)) \in \mathbb{X}_k \ominus \mathcal{B}_{2 \epsilon}$.
\end{enumerate}
\end{assumption}
The use of $2 \epsilon$ will be clarified in the proof of Theorem~\ref{th:1} of Section~\ref{subsec:theory}. Note that a terminal set $\mathbb{X}$ satisfying Assumption~\ref{ass:invset} can be defined as the set of feasible system equilibria, as in~\cite[Lemma 3]{prajapat2025safe} and~\citep{limon2018nonlinear}, or alternatively exploiting stability properties as shown in~\citep{schimperna2024robust,ravasio2024lmi,bonassi2021nonlinear} for different RNN architectures.

Considering $H \in \mathbb{N}_{>0}$ as the MPC prediction horizon, we introduce the pessimistic control action set as
\begin{equation}\label{eq:Up}
	\begin{aligned}
    \mathcal{U}^{\mathrm{p}}_k(x_k) = \{ u_{0:H-1 \mid k} \in \mathcal{U} \mid x_{0 \mid k} = x_k, x_{h+1 \mid k} = \phi(x_{h \mid k}, u_{h \mid k}), x_{h \mid k} \in \mathcal{X}^{\mathrm{p}}_k  \forall h \in [0,H-1], x_{H  \mid k} \in \mathbb{X}_k \},
	\end{aligned}    
\end{equation}
containing the input sequences ensuring that pessimistic constraints are satisfied and the state trajectory ends in the terminal set. 

We can now introduce the objective-aware safe active learning MPC problem aimed at collecting informative data while ensuring that all operational safety constraints are satisfied and without disregarding the main control objective. The following problem is solved at each time instant $k$:
\begin{subequations}
	\label{eq:safeexploMPC}
	\begin{align}
		 & \min_{u_{h \mid k}, x_{h \mid k}, \nu_{h \mid k}} \sum_{h = 0}^{H-1}  \alpha_{\nu} \cdot \nu_{h \mid k} + \mathcal{L}_{k+h}(\bar{\theta}_k^{\top} x_{h \mid k}) \label{subeq:explo_cost} \\
        & \textrm{s.t. } \,\, x_{h+1 \mid k} = \phi(x_{h \mid k}, u_{h \mid k}), \forall h \in [0,H-1], \label{subeq:explo_xdyn} \\
        & \qquad x_{0 \mid k} = x_k, \label{subeq:explo_x0} \\
        & \qquad u_{0:H-1 \mid k} \in \mathcal{U}^{\mathrm{p}}_k(x_k), \label{subeq:explo_u} \\
        & \qquad w_k(x_{h \mid k}) \geq \epsilon - \nu_{h \mid k}, \forall h \in [0,H-1], \label{subeq:explo_w} \\
        & \qquad \nu_{h \mid k} \geq 0, \forall h \in [0,H-1], \label{subeq:explo_nu}
	\end{align}
\end{subequations}
yielding the optimised exploration input sequence $u_{0:H-1 \mid k}^{\mathrm{e}}$ and the slack variable $\nu_{0:H-1 \mid k}^{\mathrm{e}}$, with $\alpha_{\nu} \in \mathbb{R}_{>0}$. Specifically, the soft constraint~\eqref{subeq:explo_w} and the slack variables~\eqref{subeq:explo_nu} are included to promote the collection of $\epsilon$-informative data, i.e., data points for which the output uncertainty $w$ exceeds the threshold $\epsilon$. As anticipated, the exact parameters $\theta^{\star}$ cannot be learned from a finite number of samples and therefore we aim to learn them up to a user-defined tolerance $\epsilon$. Moreover, constraint~\eqref{subeq:explo_xdyn} enforces the known state dynamics and~\eqref{subeq:explo_x0} ensures proper state initialisation with the measurement at time~$k$. Safety is guaranteed over the entire prediction horizon through constraint~\eqref{subeq:explo_u}, which requires the control action to belong to the pessimistic set $\mathcal{U}^{\mathrm{p}}_k(x_k)$ and, consequently, the state to belong to the pessimistic set $\mathcal{X}_k^{\mathrm{p}}$. As a result, the output physical bounds are conservatively satisfied with high probability by construction (given Corollary~\ref{corollary:beta}). Overall, the cost function~\eqref{subeq:explo_cost} minimises both the slack variables $\nu$ and the cost $\mathcal{L}$, enabling exploration of the system dynamics while simultaneously pursuing the main control objective.

Finally, for theoretical guarantees purposes, we assume that problem~\eqref{eq:safeexploMPC} employs an exact penalty for $\nu$:
\begin{assumption}
\label{ass:exactpenalty}
There exists a constant $\tilde{\alpha}_{\nu} \in \mathbb{R}_{>0} $ such that, for any $\alpha_{\nu} \geq \tilde{\alpha}_{\nu}$, the minimiser of~\eqref{eq:safeexploMPC} remains unchanged. Therefore, $\alpha_{\nu}$ in~\eqref{eq:safeexploMPC} is chosen such that $\alpha_{\nu} \geq \tilde{\alpha}_{\nu}$.
\end{assumption}

To sum up, Section~\ref{sec:exploration} addresses Objective 1 via an objective-aware safe active learning strategy~\eqref{eq:safeexploMPC} that promotes the collection of informative data for updating the output-layer parameters, ensuring that operational safety constraints are satisfied with high probability. However, the strategy could continue exploring indefinitely, thereby interfering with the main control objective and potentially degrading closed-loop performance.

\section{Goal-oriented safe active learning with finite exploration and close-to-optimal performance}
\label{sec:goalor}
In this section, we address Objective 2. In Section~\ref{subsec:pessopt}, we introduce the pessimistic and optimistic problems whose cost difference is used to determine when exploration can terminate. In Section~\ref{subsec:algo}, we present and describe the goal-oriented safe active learning algorithm, which is designed to safely learn the output model parameters only to the extent necessary to achieve close-to-optimal control performance -- hence the name \textit{goal-oriented safe active learning} -- after which it focuses exclusively on the main control objective. Finally, in Section~\ref{subsec:theory}, we state the main theoretical results for the proposed algorithm.

\subsection{Pessimistic and optimistic control problems}
\label{subsec:pessopt}
In order to understand when exploration is no longer necessary, we introduce the pessimistic and optimistic control problems, similarly to~\citep{prajapat2025safe2}. Prior to this, the following technical assumption is introduced:
\begin{assumption}
\label{ass:lips} The control cost
    $\mathcal{L}_k(y)$ is $L$-Lipschitz with respect to $y$, for all $k \in \mathbb{N}_{>0}$.
\end{assumption}

\begin{figure}
    \centering
    \includegraphics[width=0.4\linewidth]{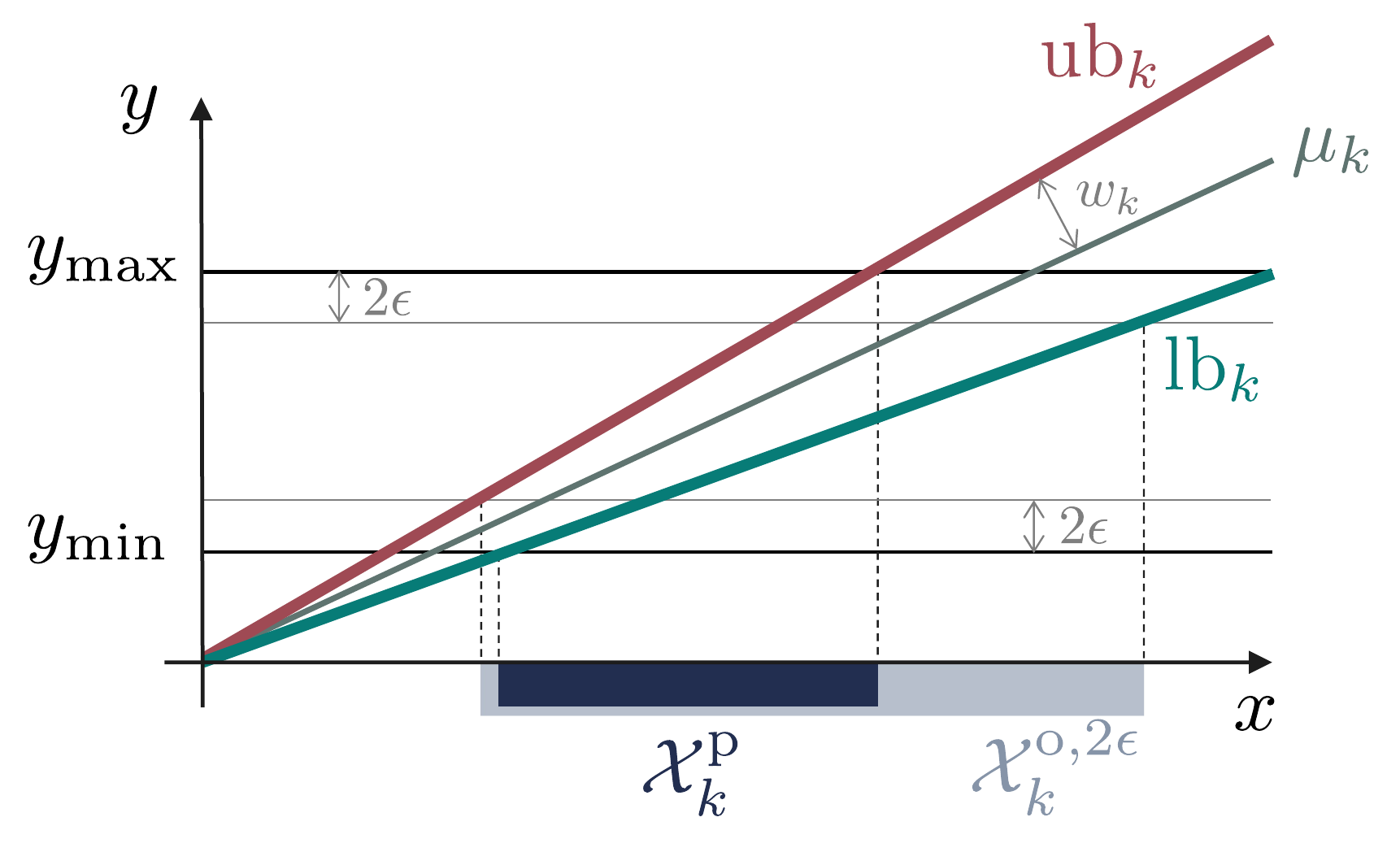}
    \caption{Schematic representation of lower (green) and upper (purple) bounds, and pessimistic (blue) and optimistic (grey) sets.}
    \label{fig:sets}
\end{figure}

Thus, the pessimistic problem is formulated as
\begin{subequations}
\label{eq:pessimisticMPC}
    \begin{align}
    J^{\mathrm{p}}_k & = \min_{u_{h \mid k}, x_{h \mid k}} \sum_{h = 0}^{H-1} \mathcal{L}_{k+h}(\bar{\theta}_k^{\top} x_{h \mid k})  + L \cdot w_k(x_{h \mid k}) \label{subeq:costpess} \\
    & \textrm{s.t. } \,\, x_{h+1 \mid k} = \phi(x_{h \mid k}, u_{h \mid k}), \forall h \in [0,H-1], \label{subeq:pess_xdyn} \\
    & \qquad x_{0 \mid k} = x_k, \label{subeq:pess_x0} \\
    & \qquad u_{0:H-1 \mid k} \in \mathcal{U}^{\mathrm{p}}_{k}(x_k), \label{subeq:upess}
	\end{align}
\end{subequations}
yielding the minimum cost $J^{\mathrm{p}}_k$ and the optimised input sequence $u^{\mathrm{p}}_{0:H-1 \mid k}$. Note that~\eqref{eq:pessimisticMPC} is equivalent to problem~\eqref{eq:safeexploMPC}, but without the exploration terms (i.e., \eqref{subeq:explo_w}-\eqref{subeq:explo_nu} and slack constraint cost in \eqref{subeq:explo_cost}), and with the additional second term in \eqref{subeq:costpess}. This term is used for guaranteeing finite exploration, as it will be clear from the proof of Theorem~\ref{th:1}. \\ 
To define an upper bound on the achievable optimal performance, we introduce the optimistic problem. First, we define the $2\epsilon$-approximation of the state and control action optimistic sets:
\begin{equation}
	\begin{aligned}
    \label{eq:Xo}
    \mathcal{X}_k^{\mathrm{o},2 \epsilon} = \{x \in \mathcal{X} \mid \operatorname{lb}_{k}(x) \leq y_{\mathrm{max}}-2 \epsilon, \operatorname{ub}_{k}(x) \geq y_{\mathrm{min}} + 2 \epsilon \},
	\end{aligned}
\end{equation}
\begin{equation}
	\begin{aligned}
    \label{eq:Uo}
    \mathcal{U}^{\mathrm{o},2 \epsilon}_k(x_k) = \{ u_{0:H-1 \mid k} \in \mathcal{U} \ominus \mathcal{B}_{2 \epsilon} \mid x_{0 \mid k}=x_k, x_{h+1 \mid k} = \phi(x_{h \mid k}, u_{h \mid k}), x_{h \mid k} \in \mathcal{X}^{\mathrm{o},2 \epsilon}_k \forall h \in [0,H-1], x_{H \mid k} \in \mathbb{X}_k \ominus \mathcal{B}_{2 \epsilon} \},
	\end{aligned}    
\end{equation}
where the latter contains the input sequences ensuring that the optimistic state constraints are satisfied and the state trajectory ends in the terminal set. The use of $2 \epsilon$ in~\eqref{eq:Xo}-\eqref{eq:Uo} is related to technical reasons necessary to guarantee finite exploration, as it will be evident in the proof of Theorem~\ref{th:1}. Figure~\ref{fig:sets} provides a schematic representation of the pessimistic and optimistic state sets. As can be seen, and considering definitions~\eqref{eq:Xkp} and~\eqref{eq:Xo}, the pessimistic set is more conservative than the optimistic one, i.e., $\mathcal{X}^{\mathrm{p}}_k \subseteq \mathcal{X}^{\mathrm{o}}_k, \forall k \in \mathbb{N}_{>0} $, with $\mathcal{X}^{\mathrm{o}}_k$ corresponding to~\eqref{eq:Xo} with $\epsilon=0$. Consequently, $\mathcal{U}^{\mathrm{p}}_k(x_k) \subseteq \mathcal{U}^{\mathrm{o}}_k(x_k), \forall k \in \mathbb{N}_{>0} $.

Moreover, we define the set
\begin{align}
    \label{eq:Thetan}
    \Theta_k = \{ \theta \in \mathbb{R}^{n_x \times n_y}  \mid  \theta^{\top}x \in [\operatorname{lb}_{k}(x), \operatorname{ub}_{k}(x)], \, \forall x \in \mathcal{X} \},
\end{align}
and thus formulate the optimistic problem as
\begin{subequations}
\label{eq:optimisticMPC}
    \begin{align}
    	J^{\mathrm{o}}_k &  = \min_{u_{h \mid k}, x_{h \mid k}, \theta}  \sum_{h = 0}^{H-1}  \mathcal{L}_{k+h}(\theta^{\top}x_{h \mid k}) \label{subeq:costopt} \\
        & \textrm{s.t. } \,\, x_{h+1 \mid k} = \phi(x_{h \mid k}, u_{h \mid k}), \forall h \in [0,H-1], \label{subeq:opt_xdyn} \\
        & \qquad x_{0 \mid k} = x_k, \label{subeq:opt_x0} \\
        & \qquad \theta \in \Theta_k, \label{subeq:thetaopt} \\ 
        & \qquad u_{0:H-1 \mid k} \in \mathcal{U}^{\mathrm{o},2 \epsilon}_{k}(x_k), \label{subeq:uopt}
	\end{align}
\end{subequations}
yielding the minimum cost $J^{\mathrm{o}}_k$. Note that the optimistic problem~\eqref{eq:optimisticMPC} achieves in principle a lower cost than the pessimistic one~\eqref{eq:pessimisticMPC}, since $\theta^{\top} x$ is allowed to vary within its lower and upper bounds~\eqref{subeq:thetaopt}, and the control variable can take values in the optimistic set~\eqref{subeq:uopt}. Conversely, in~\eqref{eq:pessimisticMPC}, the output parameters $\theta$ are fixed to the mean value computed at time $k$, i.e., $\bar{\theta}_k$, and the control action is constrained within the pessimistic set~\eqref{subeq:upess}. The latter is expected to be smaller than the $2\epsilon$-approximation of the optimistic set in~\eqref{subeq:uopt} when the output model uncertainty is large (see Figure \ref{fig:sets}).

\subsection{Algorithm description}
\label{subsec:algo}
We are now ready to introduce the overall strategy for goal-oriented safe active learning with finite exploration and close-to-optimal performance, as summarised in Algorithm~\ref{algo}.

In a nutshell, the proposed algorithm alternates between two modes. In the \textit{active exploration phase}, pessimistic control actions are applied to explore the system dynamics around the regions of optimal operation. If the difference between the pessimistic and optimistic costs, i.e., $J^{\mathrm{p}}_k-J^{\mathrm{o}}_k$, falls below a later specified threshold $\xi$, namely when the cautious and confident control actions are sufficiently close, the algorithm enters the \textit{goal-reaching phase}, as output uncertainty is sufficiently reduced. In this phase, the controller focuses exclusively on optimising the main control objective without any further exploration, while enforcing pessimistic constraints to ensure safety. If, at any time, the difference between the pessimistic and optimistic costs exceeds the threshold $\xi$, the algorithm returns to the exploration phase. This may occur, for instance, if the system begins to operate in an unexplored region characterised by high uncertainty due to variations in the cost function (e.g., a change in the tracking reference). 

\renewcommand{\alglinenumber}[1]{#1:}
\begin{algorithm}[t!]
\caption{Goal-oriented safe active learning with finite exploration and close-to-optimal performance}
	\label{algo}
	\begin{algorithmic}[1] 
		\State Initialise $x_1 \in \mathbb{X}_1$, $\Lambda^{-1}_0$, $\bar{\theta}_0 $, $k=1$, $k_0=1$, $n=0$, $\epsilon$

		\State \textbf{while True}:

        \State \quad \text{Measure} $x_k$, $\tilde{y}_k^{\star}$ from~\eqref{eq:sys}
        
        \State \quad \text{Update} $\Lambda^{-1}_k$, $\bar{\theta}_k$ with~\eqref{eq:update_param1}, and $\mu_k$, $\Sigma^2_k$ with~\eqref{eq:update_y}
                
        \State \quad $J^{\mathrm{p}}_k$, $u^{\mathrm{p}}_{0:H-1 \mid k}$ $\gets$ Solve~\eqref{eq:pessimisticMPC} with $x_k$, $\Lambda_k^{-1}$, $\bar{\theta}_k$ 
        
        \State \quad $J^{\mathrm{o}}_k$ $\gets$ Solve~\eqref{eq:optimisticMPC} with $x_k$, $\Lambda_k^{-1}$, $\bar{\theta}_k$ 
        
		\State \quad \textbf{if} $J^{\mathrm{p}}_k-J^{\mathrm{o}}_k > \xi$ \textbf{then} \quad (\texttt{Active exploration})
        
        \State \quad \quad $n \gets n+1$
             
        \State \quad \quad $u^{\mathrm{e}}_{0:H-1 \mid k}$,  $\nu^{\mathrm{e}}_{0:H-1 \mid k}$  $\gets$ Solve~\eqref{eq:safeexploMPC}  with $x_k$, $\Lambda_k^{-1}$, $\bar{\theta}_k$ 
                
        \State \quad \quad $u_{k} \gets u^{\mathrm{e}}_{0 \mid k}$
        
        \State \quad \quad \textbf{if} $\exists h \in [1, H-1] \mid \nu^{\mathrm{e}}_{h \mid k} = 0 $ \textbf{then}
        
        \State \quad \quad \quad $h^{\star} \gets \min \{h \in [1, H-1] \mid \nu^{\mathrm{e}}_{h \mid k} = 0 \}$

        \State \quad \quad \textbf{else}

        \State \quad \quad \quad $h^{\star} = H $
        
        \State \quad \quad $k_n \gets k+h^{\star}$
        
        \State \quad \quad $k \gets k+1$

        \State \quad \quad \textbf{for} $h = 1,\hdots,h^{\star}-1$ 
                
        \State \quad \quad \quad $u_{k} \gets u^{\mathrm{e}}_{h \mid k-h}$

        \State \quad \quad \quad \text{Measure} $x_k$, $\tilde{y}_k^{\star}$ from~\eqref{eq:sys}
        
        \State \quad \quad \quad \text{Update} $\Lambda^{-1}_k$, $\bar{\theta}_k$ with~\eqref{eq:update_param1}, and $\mu_k$, $\Sigma^2_k$ with~\eqref{eq:update_y}

        \State \quad \quad \quad $k \gets k+1$
        
        \State \quad \textbf{else} \quad (\texttt{Goal-reaching})
        
        \State \quad \quad $n=0$
        
        \State \quad \quad $k_n \gets k+1$
                                
        \State \quad \quad $u_k \gets u^{\mathrm{p}}_{0 \mid k}$

        \State \quad \quad $k \gets k+1$
        
	\end{algorithmic}
\end{algorithm} 

In detail, the algorithm begins with an initialisation phase, where the initial states and parameter priors are set, along with the discrete-time index $k$, the time index at which an $\epsilon$-informative state is collected $k_n$, the counter $n$ that tracks the number of iterations in each exploration phase, and the user-defined learning tolerance $\epsilon$ (line 1). The algorithm proceeds iteratively for each simulation step $k = 0,1,\hdots$ (line 2). First, the system measurements at time $k$ are collected (line 3), and the mean and covariance of both the parameters and the output are updated accordingly (line 4). Then, the pessimistic~\eqref{eq:pessimisticMPC} and optimistic~\eqref{eq:optimisticMPC} problems are solved (lines 5-6) to check whether the difference between their costs exceeds the switching threshold $\xi$ (line 7). Note that~\eqref{eq:pessimisticMPC} and~\eqref{eq:optimisticMPC} may be solved in parallel. \\
If the condition in line 7 is satisfied, then the optimistic and pessimistic costs are significantly different, indicating that exploration is required: the exploration counter $n$ is incremented (line 8). During this phase, the objective-aware safe exploration problem~\eqref{eq:safeexploMPC} is solved (line 9) and the first control action $u_{0 \mid k}^{\mathrm{e}}$, corresponding to the just updated model, is applied to the system (line 10). Next, $h^{\star}$ is determined as the first time instant $h \in [1,H-1]$ at which an $\epsilon$-informative state is predicted to occur, i.e., when~\eqref{subeq:explo_w} is satisfied with zero slack (lines 11-12). While lines 13-14 are included for technical completeness, they are never executed under a specific choice of $\xi$. This choice guarantees that if the condition in line 11 is not satisfied, the one in line 7 will not be satisfied either, as will be evident from the proof of Theorem~\ref{th:1}. Then, the time index corresponding to this $\epsilon$-informative state is stored (line 15), and $k$ is incremented (line 16). Subsequently, the optimal control sequence is applied until the $\epsilon$-informative state is reached, safely controlling and exploring the system dynamics (lines 17-18). Measurements are collected (line 19), and the mean and covariance of both the parameters and the output are updated recursively (line 20), after which $k$ is incremented (line 21). \\
On the other hand, if the difference between the pessimistic and optimistic costs falls below the switching threshold $\xi$ (line 22), the cautious (pessimistic) and the confident (optimistic) solutions nearly coincide, meaning that the system has been sufficiently explored in that region of operation. In this case, the counter of exploration iterations $n$ is reset to zero (line 23), $k_n$ is updated accordingly (line 24), and the first control action of the pessimistic sequence is applied (line 25) to steer the system toward the control objective while ensuring safety. The time step $k$ is finally incremented (line 26).

\subsection{Theoretical guarantees}
\label{subsec:theory}
Before introducing the theorem providing theoretical guarantees for Algorithm~\ref{algo}, we define the $2\epsilon$-approximation of the real state and control action sets as
\begin{equation}
	\begin{aligned}
    \mathcal{X}^{\star,2\epsilon} & = \{x \in \mathcal{X} \:  \mid  \: y^{\star} = \theta^{\star \top} x \in \mathcal{Y} \ominus \mathcal{B}_{2\epsilon} \},
	\end{aligned}
\end{equation}
\begin{equation}
	\begin{aligned}
    \mathcal{U}^{\star,2\epsilon}_k(x_k) = \{ u_{0:H-1 \mid k} \in \mathcal{U} \ominus \mathcal{B}_{2\epsilon}  \mid  x_{0 \mid k}=x_k, x_{h+1 \mid k} = \phi(x_{h \mid k}, u_{h \mid k}), x_{h \mid k} \in \mathcal{X}^{\star,2\epsilon}  \forall h \in [0,H-1], x_{H \mid k} \in \mathbb{X}_k \ominus \mathcal{B}_{2\epsilon}\}.
	\end{aligned}    
\end{equation}

We are now ready to state our main theoretical result, whose proof is provided in Appendix~\ref{appendix:prooftheo1}.

\begin{theorem}
\label{th:1}
    Let Assumptions~\ref{ass:prior}-\ref{ass:lips} hold, and suppose the initial condition satisfies $x_1 \in \mathbb{X}_1$. With at least $1-\delta$ probability, the following properties hold:
    \begin{enumerate}
        \item recursive feasibility: all optimisation problems in Algorithm~\ref{algo} are feasible $\forall k \in \mathbb{N}_{>0}$;
        
        \item safety: Algorithm~\ref{algo} guarantees safety for the unknown system~\eqref{eq:sys} at all times, i.e., $x_k \in \mathcal{X}$, $u_k \in \mathcal{U}$, $y_k^{\star} \in \mathcal{Y} = [y_{\mathrm{min}}, y_{\mathrm{max}}]$, $\forall k \in \mathbb{N}_{>0}$;
        
        \item finite termination of exploration: each exploration phase in Algorithm~\ref{algo} terminates in at most $n^{\star} \in \mathbb{N}$ iterations guaranteeing a switch to the goal-reaching phase, i.e., $J^{\mathrm{p}}_{k_{n^{\star}}}-J^{\mathrm{o}}_{k_{n^{\star}}} \leq \xi$, with $\xi = 2 L \epsilon H$;
        
        \item close-to-optimal performance: if $J^{\mathrm{p}}_k-J^{\mathrm{o}}_k \leq \xi$, Algorithm~\ref{algo} returns a control sequence $u_{0:H-1 \mid k}^{\mathrm{p}}$ that achieves close-to-optimal performance, i.e.,
        \begin{equation}
        \begin{aligned}
        \label{eq:closetoopt}
            J_k(x_k, \theta^{\star}, u_{0:H-1 \mid k}^{\mathrm{p}})  \leq \min_{u_{0:H-1 \mid k}\in \mathcal{U}^{\star,2 \epsilon}_k(x_k)}  J_k(x_k, \theta^{\star}, u_{0:H-1 \mid k}) + \xi
        \end{aligned}
        \end{equation}
        where $J_k(x_k, \theta, u_{0:H-1 \mid k}) = \sum\limits_{h=0}^{H-1} \mathcal{L}_{k+h}(\theta^{\top}x_{h \mid k})$, with $x_{h+1 \mid k}=\phi(x_{h \mid k}, u_{h \mid k})$, for $h \in [0,H-1], x_{0 \mid k}=x_k$. 
    \end{enumerate}
\end{theorem}

To sum up, Section~\ref{sec:goalor} addresses Objective 2 via a goal-oriented safe active learning algorithm that, in addition to learning the output-layer parameters with feasibility and safety guarantees, does so only to the extent needed to achieve control performance comparable to that of an omniscient MPC, thereby terminating exploration when it is no longer needed. Specifically, by selecting a switching threshold $\xi$ proportional to the confidence level $\epsilon$, the algorithm terminates exploration in a finite number of iterations and returns a pessimistic input sequence that achieves a cost close to that attainable by knowing $\theta^{\star}$ and using the 2$\epsilon$-approximation of the real input constraint set, i.e., \eqref{eq:closetoopt}.

Overall, the advantage of this strategy, based on the pessimistic and optimistic problems, can be inferred from the theorem. In particular, the pessimistic problem is formulated to guarantee safety despite uncertainty, whereas the optimistic problem is introduced to determine when exploration can be terminated, namely when the output parameters have been learned sufficiently such that adopting either a cautious or a confident approach with respect to the operational safety constraints leads to nearly the same cost.

\section{Case study}
\label{sec:casestudy}
In this section, we evaluate the proposed framework on a benchmark energy system. The system is described in Section~\ref{subsec:setting}, the control strategy adopted to optimise system operation is presented in Section~\ref{subsec:conroldesign}, and the numerical results are reported in Section~\ref{subsec:numres}.

\subsection{Settings}
\label{subsec:setting}
\begin{figure}
    \centering
    \includegraphics[width=0.4\linewidth]{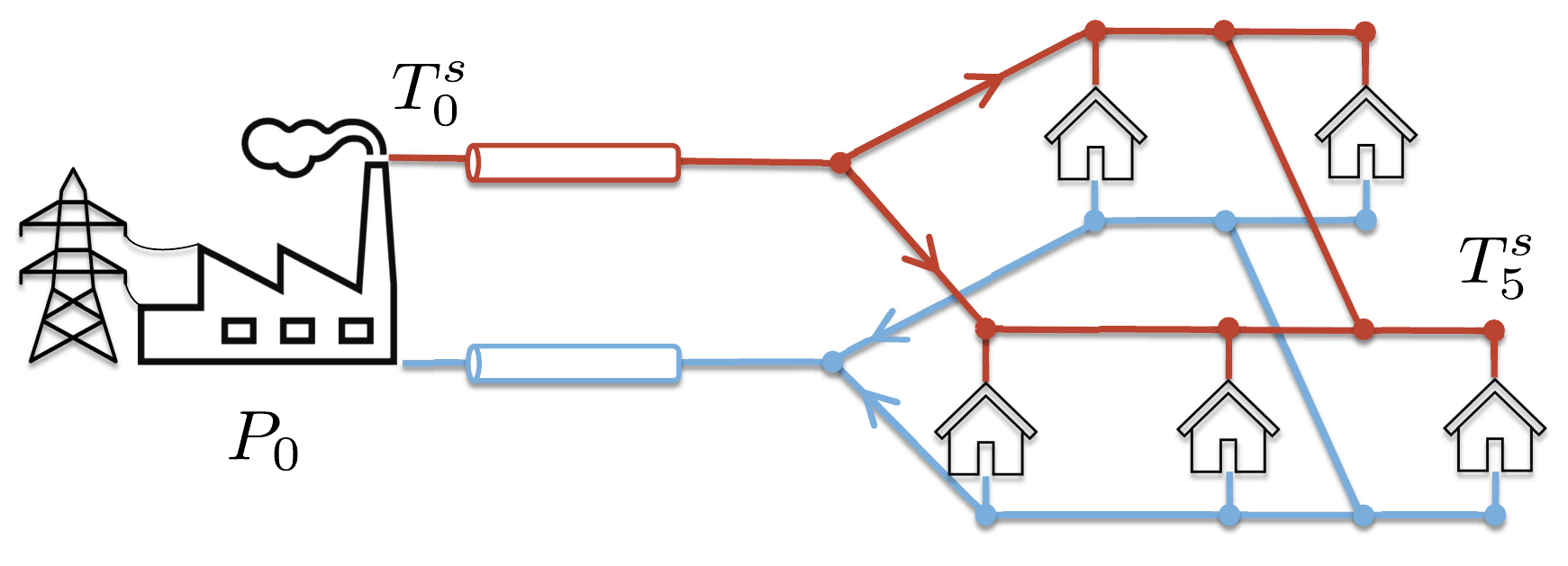}
    \caption{Schematic representation of the AROMA DHS, showing the heating station, the supply and return pipelines, and the five thermal loads, together with the considered variables~\citep{de2024physics}.}
    \label{fig:aroma}
\end{figure}

The proposed framework is tested in simulation on a district heating system (DHS), an energy distribution network crucial for achieving decarbonisation targets. A DHS typically consists of a central heating station with multiple thermal generators and a network of insulated water pipelines that deliver heat to thermal loads. These loads use local heat exchangers to absorb the delivered heat for indoor heating and domestic hot water~\citep{la2023optimal}. The specific case study considered in this work is the AROMA DHS, presented in~\citep{krug2021nonlinear} and depicted in Figure~\ref{fig:aroma}. Its physical model is leveraged to develop a dynamic simulator within a dedicated Modelica library~\citep{nigro2024control}, which serves as a digital twin to generate input-output data for training the data-based model. In this work, we consider as control input the supply temperature at the heating station, denoted by \mbox{$T_0^s$}, and as controlled outputs the power produced by the heating station $P_0$ and the supply temperature of the most distant load with respect to the heating station, i.e., $T_5^s$, since it experiences the largest thermal losses and is therefore the most critical in terms of constraints satisfaction~\citep{la2023optimal}. For the sake of clarity, the five load thermal demands are assumed to be constant.

The baseline model~\eqref{eq:sys} for the AROMA DHS is identified using a gated recurrent unit (GRU) network, belonging to the RNN family, due to its strong approximation capability and relatively simple architecture~\citep{bonassi2022recurrent}. The GRU model is formulated as
\begin{subequations}
	\label{eq:gru}
	\begin{numcases}{}
    \begin{aligned}
    \label{eq:x_gru}
		x_{k+1} = z_k \circ x_{k} + (1- z_k) \circ  \tanh (W_r  u_k  + U_r  f_k  \circ x_k + b_r)
    \end{aligned}\\
    \begin{aligned}
    \label{eq:y_gru}
		y_k^{\star} = U_o^{\star} x_k + b_o^{\star} = \theta^{\star \top} [x_k \: \, 1]
    \end{aligned}
	\end{numcases}
\end{subequations}
where the state equation~\eqref{eq:x_gru} corresponds to~\eqref{eq:x_sys}, the output equation~\eqref{eq:y_gru} to~\eqref{eq:y_sys}, and the update and forget gates, $z_k$ and $f_k$ respectively, are defined as
\begin{equation}
	\label{eq:grugates}
	\begin{aligned}
		& z_{k} = f_{\sigma}(W_z u_k + U_z x_k + b_z), \\
		& f_{k} = f_{\sigma}(W_f u_k+ U_f x_k + b_f),
	\end{aligned}
\end{equation}
with $f_{\sigma}$ denoting the sigmoid activation function. The baseline model is trained offline for 200 epochs using the library developed in~\citep{ssnet} with a 10-day input-output dataset collected from the physical simulator at a sampling time of 5 minutes. It features an input layer for $T_0^s$, a single hidden layer with six states, and two output layers corresponding to the system outputs $T_5^s$ and $P_0$. Overall, $x \in \mathbb{R}^{6}$, $u \in \mathbb{R}^{1}$, and $y \in \mathbb{R}^{2}$.

\subsection{Control design}
\label{subsec:conroldesign}
To optimally control the AROMA DHS, we implement an MPC scheme, whose cost function is defined as
\begin{align}
     \min_{T_{0, h \mid k}^s} \mathcal{L} = \sum_{h = 0}^{H-1}  (c_{k+h}^{el} \cdot P_{0,h \mid k} \cdot \tau ) +  c_T \cdot \lvert T_{5,H \mid k}^{s}-T^{\star} \rvert. \label{subeq:costf}
\end{align}
In detail, the first term of the cost function aims to minimise the production cost of the heating station, where $c^{el}$ is the electricity price reported in Figure~\ref{fig:omniscientmpc}(a) and $\tau = 5$ min is the sampling time. The second term of the cost function is a terminal cost, weighted by $c_T$, which penalises deviations of the final load supply temperature from a nominal reference value $T^{\star} = 80^{\circ}\mathrm{C}$. In addition, the MPC contains constraints enforcing operational lower and upper limits on the input temperature and on the output temperature and power, as well constraints enforcing the dynamical model of the system~\eqref{eq:gru}. The complete MPC formulation is provided in~\citep{de2024physics}. 

We first evaluate this MPC in an omniscient scenario, i.e., assuming a predictive model with the parameters identified offline and known exactly ($\theta^{\star}$). \\
However, as discussed in Section~\ref{sec:intro}, access to such an omniscient predictive model is uncommon, particularly for large-scale and complex systems such as DHSs. We therefore then evaluate the learning-based algorithm proposed in Section~\ref{subsec:algo}, in which the MPC operates with a model that is learned online. Specifically, following the Bayesian last-layer approach presented in Section~\ref{subsec:bll}, the hidden-layer parameters $\{W_r, U_r, b_r, W_z, U_z, b_z, W_f, U_f, b_f\}$ are assumed to be known and fixed, having been identified during the offline training procedure (i.e., from the omniscient model). In contrast, the output-layer parameters $\theta = [U_o \: \, b_o]^{\top}$, with $U_o \in \mathbb{R}^{2 \times 6}$ and $b_o \in \mathbb{R}^{2 \times 1}$, are learned over time through the proposed goal-oriented safe active learning strategy.

\subsection{Numerical results}
\label{subsec:numres}
Both the omniscient and the proposed learning-based MPC are tested over a one-day simulation, with a sampling time $\tau=5$ min and a prediction horizon $H=24$. \\
In addition, for the proposed learning-based MPC, the noise variance is set to $\sigma^2=0.001$, the uncertainty threshold is chosen as $\epsilon=5\cdot\sigma$, the probability level is $\delta = 0.01$, and the initial guesses are selected as $\theta_0 = 0.3 \cdot \theta^{\star}$, $\Lambda_0 = 0.3 \cdot I_{(n_x+1) \cdot n_y}$. The invariant set is computed similarly to~\citep{limon2018nonlinear}. \\
All computations are performed on a laptop equipped with an Intel Core i7-10750H processor, using Python 3.12, the CasADi environment, and the Ipopt solver.

We compare the omniscient MPC and the proposed learning-based MPC with a rule-based strategy that is typically used to control real-world DHSs. In this strategy, the control input is kept constant at $\overline{T}_0^s = 80^{\circ}\mathrm{C}$~\citep{la2023optimal}. To quantitatively evaluate this and the other control strategies, we compute the daily production cost as \mbox{$C_p=\sum\limits_{k=1}^T c_k^{el} P_{0,k}^{\star} \tau$}, where $P_0^{\star}$ is the effective power produced by the heating station and $T=24$ h$/\tau=288$. As reported in Table~\ref{table:costcomparison}, adopting this rule-based strategy yields a daily production cost of $C_p = 7458.89$ \euro.

\begin{figure}[t!]
	\centering
	\captionsetup[subfloat]{labelfont=scriptsize,textfont=scriptsize}
	\subfloat[]{\includegraphics[width=0.3\textwidth]{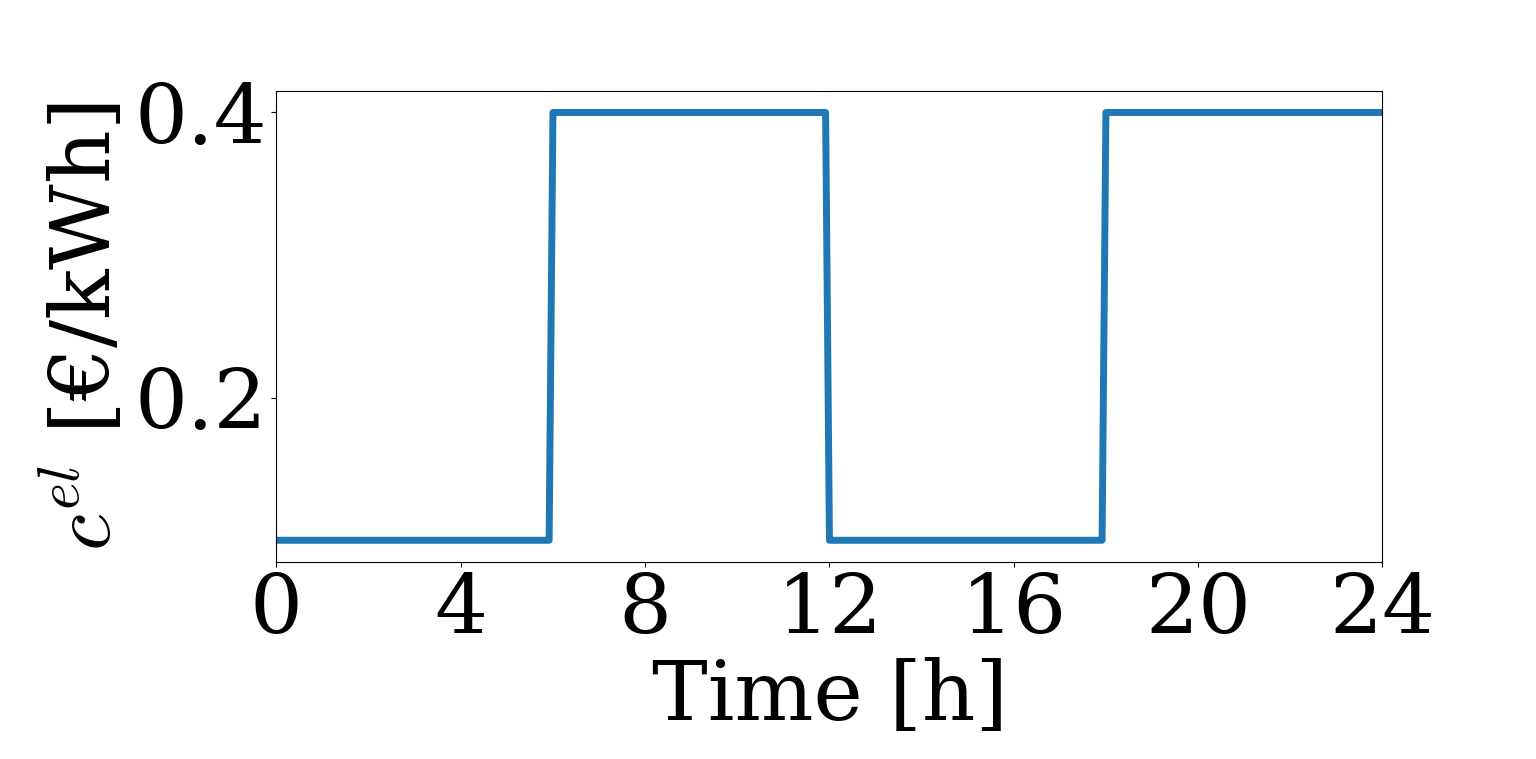} } 
	\subfloat[]{\includegraphics[width=0.3\textwidth]{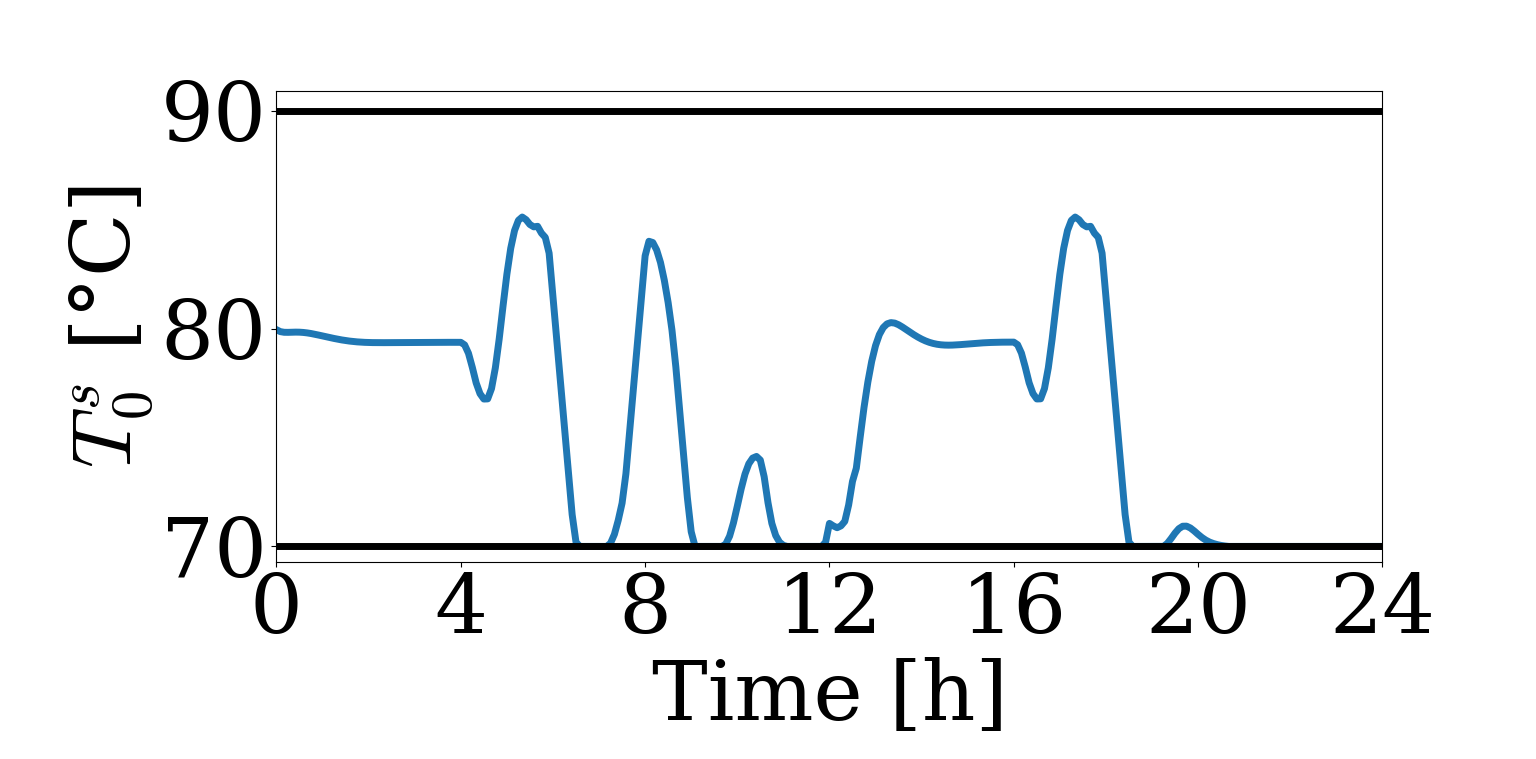} } \\
	\subfloat[]{ \includegraphics[width=0.3\textwidth]{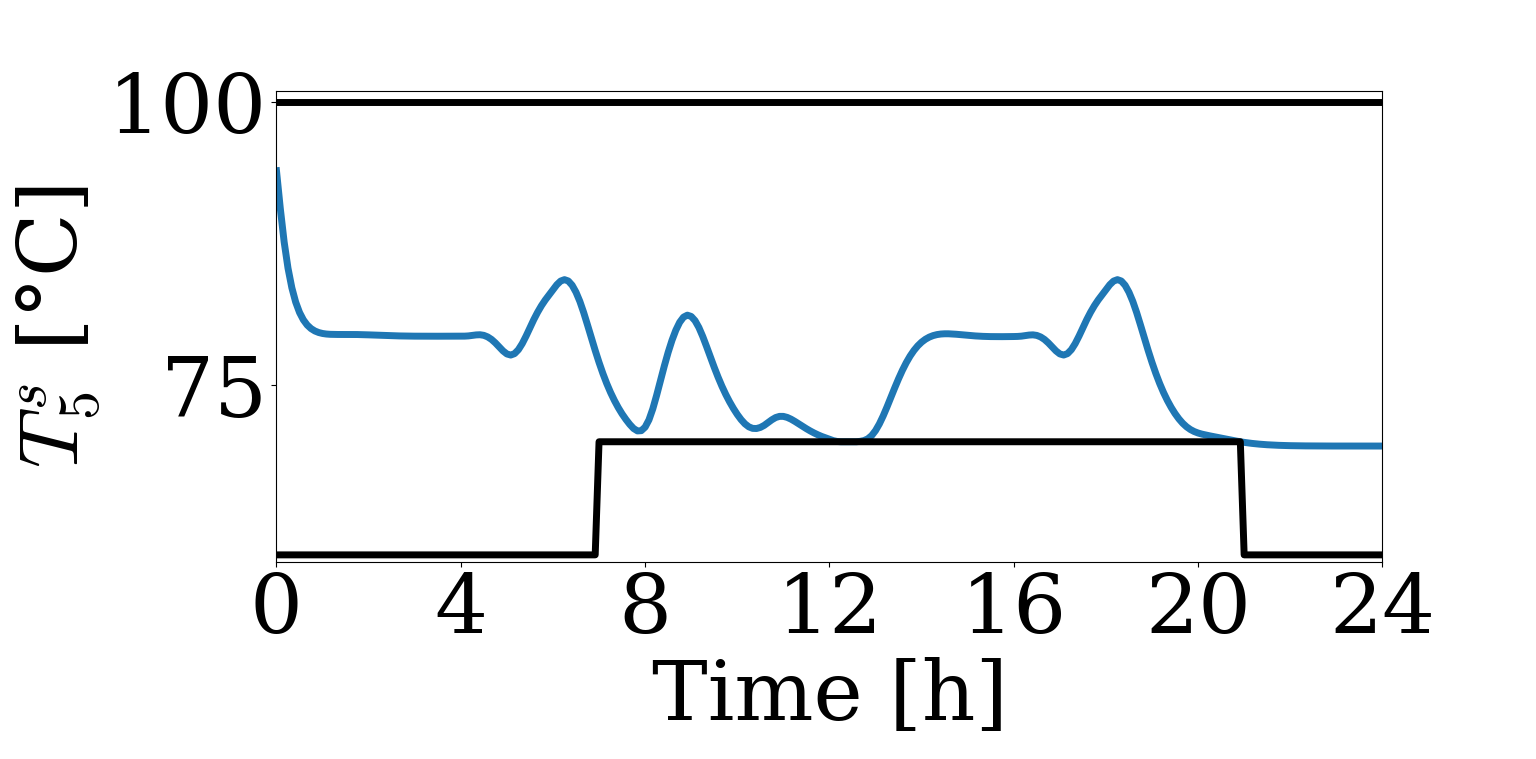} } 
	\subfloat[]{ \includegraphics[width=0.3\textwidth]{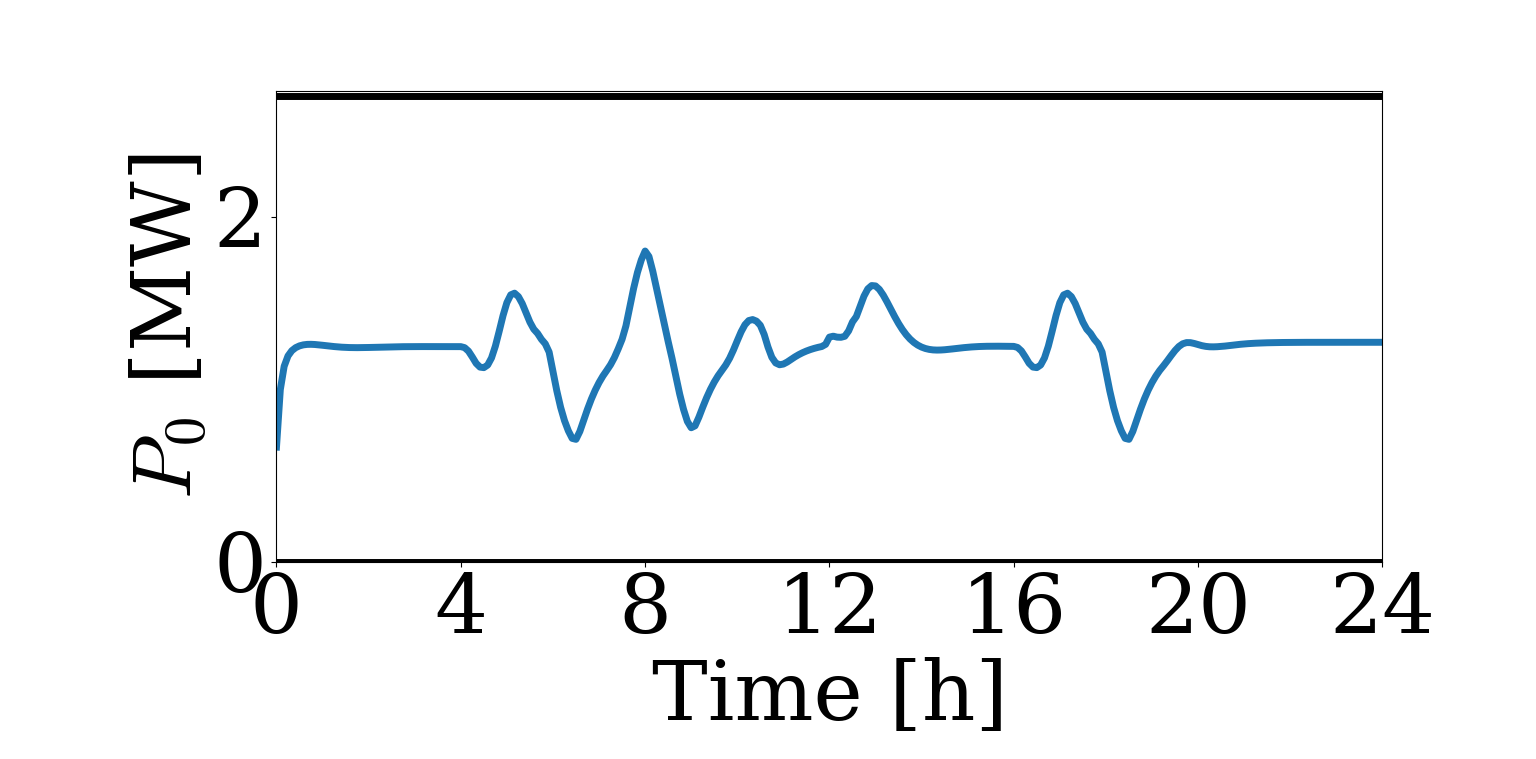} }
	\caption{Simulation results of the omniscient MPC. (a) Electricity price. (b) Optimised control input (blue) and corresponding constraints (black). (c) Load supply temperature (blue) and corresponding constraints (black). (d) Plant power (blue) and corresponding constraints (black). }
	\label{fig:omniscientmpc}
\end{figure}

We then evaluate the performance of the omniscient MPC, which has an average solving time of $t_{\mathrm{avg}}=0.4$s. This optimisation-based strategy achieves cost savings compared to the rule-based approach. As reported in Table~\ref{table:costcomparison}, it results in a daily production cost of $C_p = 7199.90$~\euro, corresponding to a $3.4 \%$ reduction compared to the rule-based strategy. The optimised control input is reported in Figure~\ref{fig:omniscientmpc}(b), showing that when the electricity price is high, the MPC lowers the supply temperature of the heating station, whereas $T_0^s$ is increased when the electricity price decreases. The same behaviour can be observed in the load supply temperature and plant power, depicted in Figures~\ref{fig:omniscientmpc}(c) and~\ref{fig:omniscientmpc}(d), respectively. Overall, the predictive capability of the MPC enables a reduction in production costs by charging the DHS when it is economically advantageous and discharging it otherwise. 

\begin{figure}[t!]
	\centering
	\captionsetup[subfloat]{labelfont=scriptsize,textfont=scriptsize}
	\subfloat[]{\includegraphics[width=0.3\textwidth]{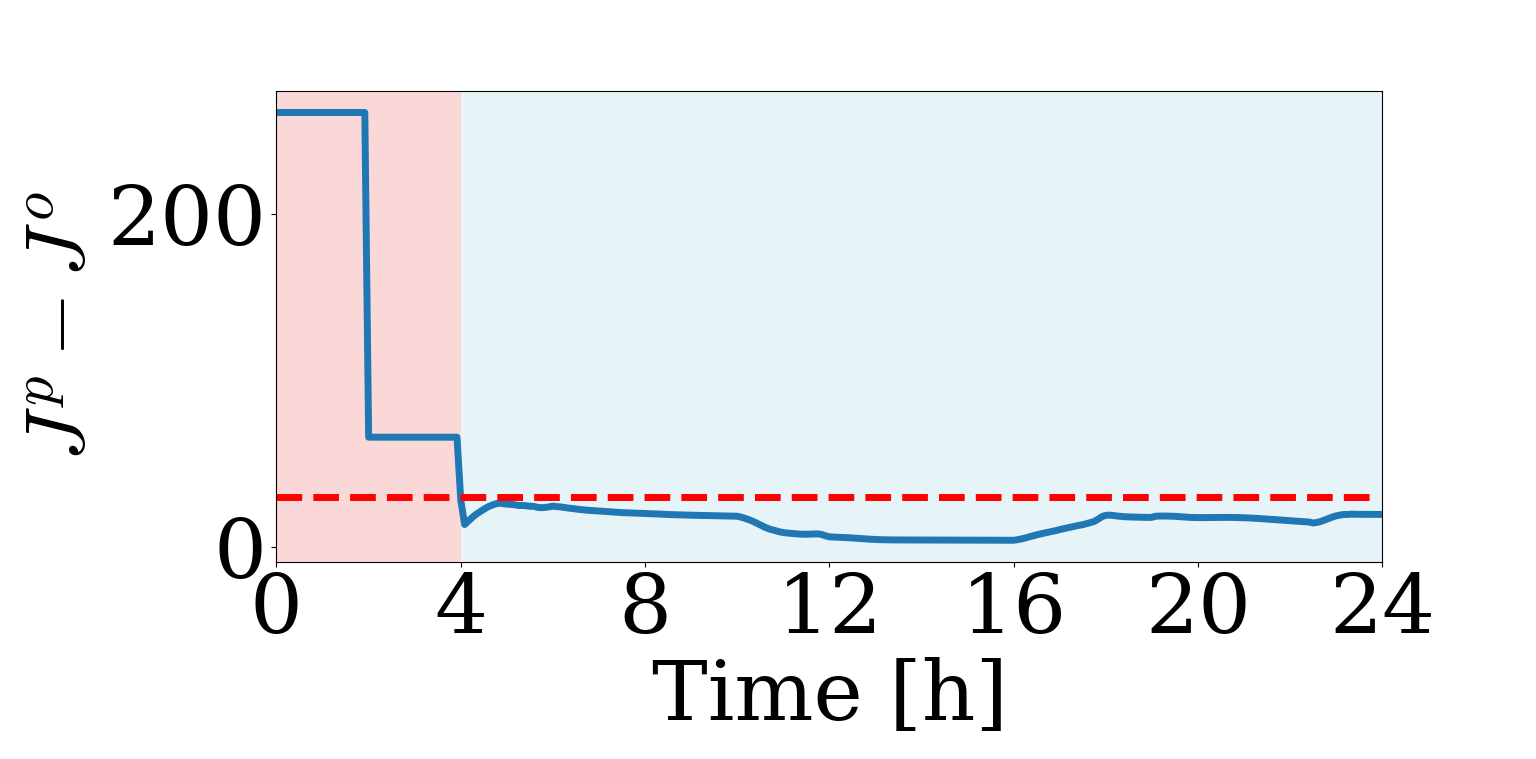} } 
	\subfloat[]{\includegraphics[width=0.3\textwidth]{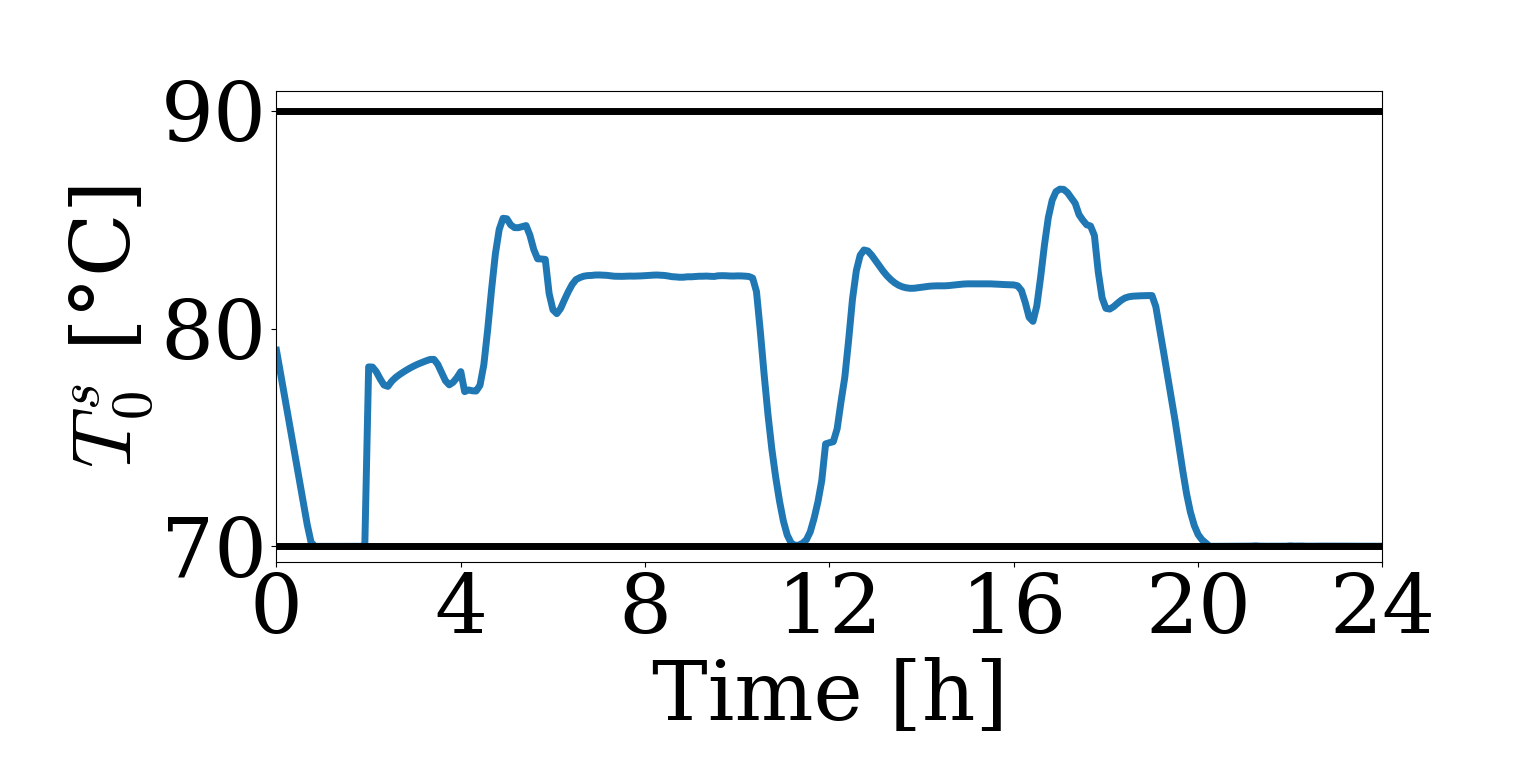} } 
	\subfloat[]{\includegraphics[width=0.3\textwidth]{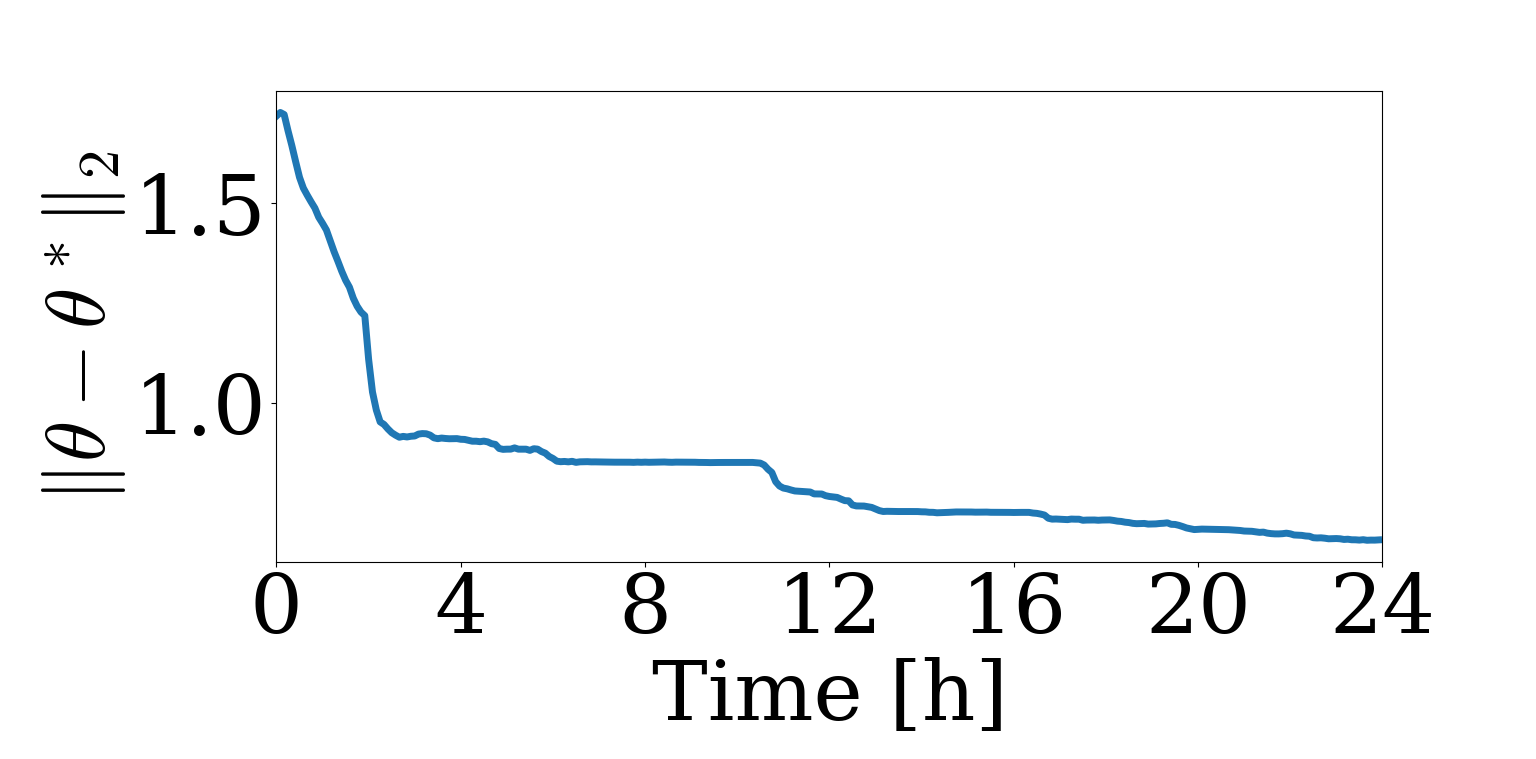} } \\
	\subfloat[]{ \includegraphics[width=0.3\textwidth]{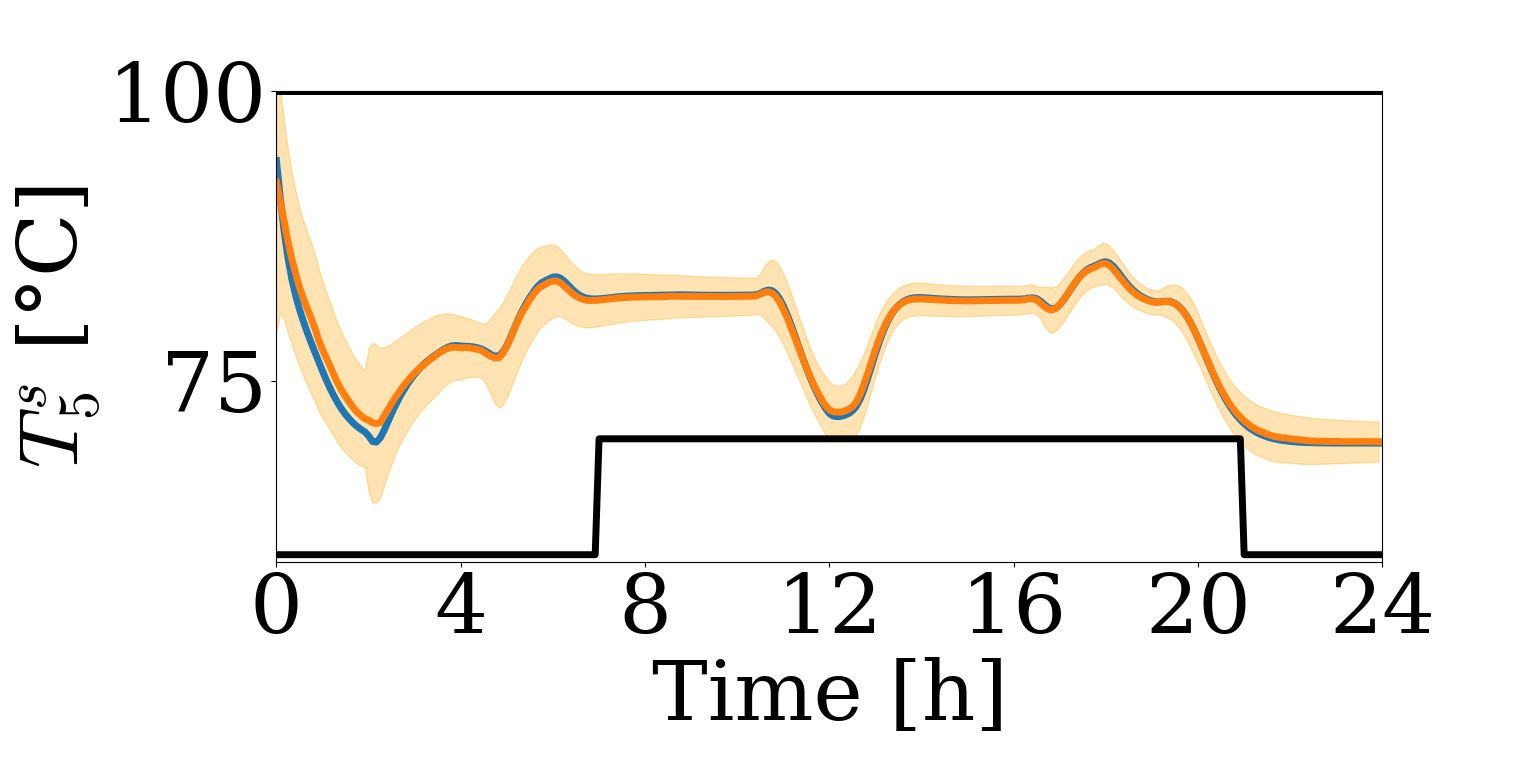} } 
	\subfloat[]{ \includegraphics[width=0.3\textwidth]{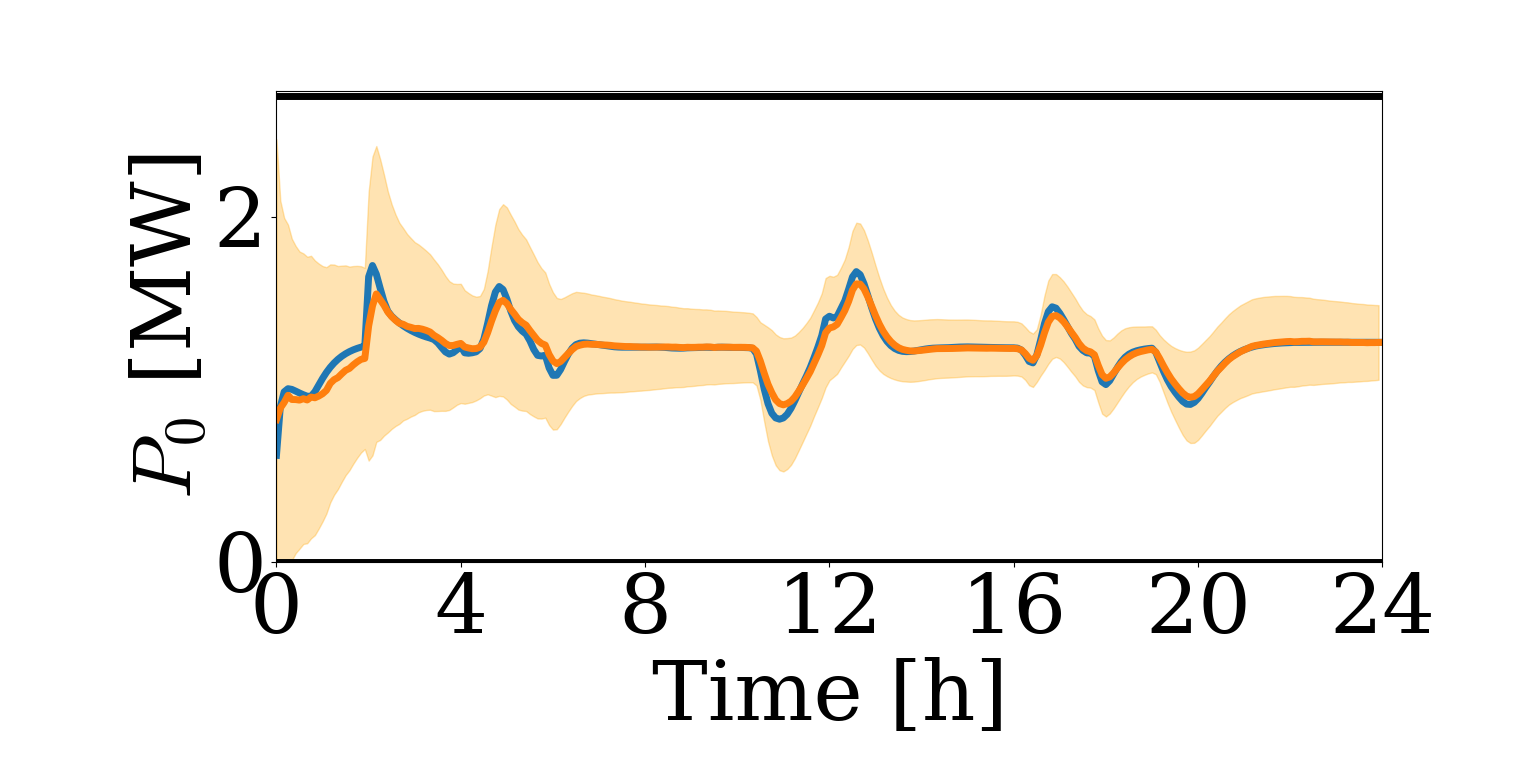} }
	\caption{Simulation results of the proposed learning-based MPC. (a) Cost difference $J^{\mathrm{p}}-J^{\mathrm{o}}$ (blue) and switching threshold $\xi$ (dashed red); background colours depict phases (exploration in pink, goal-reaching in light-blue). (b) Optimised control input (blue) and corresponding constraints (black). (c) Evolution of the 2-norm of the parameters error. (d) Load supply temperature trend comparing the ground truth (blue) and the learned one with its confidence intervals (orange). Constraints are displayed in black. (e) Power trend comparing the ground truth (blue) and the learned one with its confidence intervals (orange). Constraints are displayed in black. }
	\label{fig:exploration}
\end{figure}

Finally, we evaluate the performance of the proposed learning-based MPC, which has an average solving time of $t_{\mathrm{avg}}=1.6$s. The simulation results are presented in Figure~\ref{fig:exploration}. Figure~\ref{fig:exploration}(a) shows, in blue, the difference $J^{\mathrm{p}}-J^{\mathrm{o}}$, together with the switching threshold $\xi$ indicated by a red dashed line. The same plot also displays the phase of the algorithm: a pink background corresponds to the exploration phase, whereas a light-blue background represents the goal-reaching phase. It can be observed that the algorithm remains in exploration mode until approximately 4 a.m.; afterwards, since $J^{\mathrm{p}}-J^{\mathrm{o}}<\xi$, it switches to the goal-reaching phase, thereby demonstrating finite-time termination of exploration. Figure~\ref{fig:exploration}(b) depicts the control input $T_0^s$, whose profile is consistent with that of the omniscient MPC shown in Figure~\ref{fig:omniscientmpc}(b), with noticeable differences due to the initial exploration phase and, subsequently, to the tightened constraints imposed during pessimistic goal-reaching. The optimised input allows the model parameters to be progressively refined over time, as illustrated in Figure~\ref{fig:exploration}(c), where the estimation error $\lVert \theta-\theta^{\star}\rVert_2$ gradually decreases across iterations. Note that the error does not converge to zero, since the real parameters cannot be learned from a finite number of samples, but can only be reduced such that the output uncertainty falls below the prescribed threshold $\epsilon$. This results in a progressive alignment between the learned output (orange) and the ground-truth output (blue), as shown in Figures~\ref{fig:exploration}(d) and~\ref{fig:exploration}(e) for the load supply temperature and the plant power, respectively. From these plots, it can also be observed that, as stated in Corollary~\ref{corollary:beta}, the actual output always remains within the lower and upper bounds (shaded orange) and therefore consistently satisfies the operational safety constraints depicted in black. This behaviour is particularly evident under critical operating conditions, such as when the lower bound on the load supply temperature increases from $60^{\circ}\mathrm{C}$ to $70^{\circ}\mathrm{C}$ between $7$ a.m. and $9$ p.m., in agreement with the daily thermal demand profile~\citep{la2023optimal}. The close-to-optimal performance stated in Theorem~\ref{th:1} is confirmed by the daily production cost achieved by this control strategy, which amounts to $C_p = 7207.62$ \euro, corresponding to a $3.3 \%$ reduction compared with the rule-based strategy. As expected, this cost is slightly higher than that obtained with the omniscient MPC, since the algorithm initially operates in exploration mode, exciting the system and temporarily prioritising model learning over production cost minimisation. In addition, during goal-reaching, the constraints are tighter than those of the omniscient MPC due to the adoption of pessimistic constraints. 

In summary, the proposed framework enables learning of the model parameters during system operation while simultaneously exploring the system dynamics and optimising the main control objective, always respecting operational safety constraints (Objective 1). When the model parameters have been learned to achieve sufficiently good control performance, the exploration phase terminates and, thereafter, the resulting solution is close to optimal (Objective 2), yielding a production cost comparable to that obtained with the omniscient model and still achieving a promising reduction compared to the rule-based strategy commonly adopted in DHS plants. 

\begin{table}[t!]
	\centering
	\caption{Comparison of daily production costs achieved by the different control strategies.}
	\begin{tabular}{l!{\vrule}c}
        Control strategy & Cost [\euro/day] \\ [0.1cm] \hline \\[-3px]
		Rule-based strategy & 7458.89 \\ [2px] 
		Omniscient MPC & 7199.90 \\ [2px] 
		Proposed learning-based MPC & 7207.62
	\end{tabular}
	\label{table:costcomparison}
\end{table}

\section{Conclusion}
\label{sec:conclusions}
This work proposes a goal-oriented safe active learning algorithm for the online adaptation of a Bayesian last-layer recurrent neural network (RNN) embedded within a model predictive controller (MPC). The algorithm alternates between an exploration and a goal-reaching phase, while always ensuring satisfaction of operational safety constraints. During the objective-aware exploration phase, the MPC explores the system dynamics to collect informative data for the recursive update of the RNN output-layer parameters while also focusing on the main control objective. When the costs associated with a pessimistic (i.e., cautious) and an optimistic (i.e., confident) MPC formulation become sufficiently close, the algorithm switches to the goal-reaching phase, where it focuses exclusively on the main control objective. The method enables computationally efficient online parameter adaptation while guaranteeing recursive feasibility, safety, and close-to-optimal controller performance 
within finite time. The framework is validated on a benchmark district heating system, demonstrating that model predictions are progressively refined over time while operational safety constraints are always respected. Moreover, the algorithm automatically transitions from exploration to goal-reaching, achieving economic cost savings comparable to those obtained by an omniscient MPC. \\
Future work will focus on integrating the proposed framework with control learning, in which safe exploration of the system dynamics is used not only to adapt the model but also to learn control parameters, e.g., cost function weights.

\section*{Acknowledgments}
This work was supported by the Swiss National Science Foundation as a part of NCCR Automation, a National Centre of Competence in Research, (grant number 51NF40\_225155), and by the Next-Generation EU (Italian PNRR - M4 C2, Invest 1.3 - D.D. 1551.11-10-2022, PE00000004, CUP MICS D43C22003120001) as a part of the MICS (Made in Italy - Circular and Sustainable) Extended Partnership.
Manish Prajapat is supported by ETH AI center doctoral fellowship.
Anna Scampicchio acknowledges that this work was partially supported by the Wallenberg AI, Autonomous Systems and
Software Program (WASP) funded by the Knut and Alice Wallenberg Foundation.

\bibliographystyle{agsm}  
\bibliography{Bibliography.bib}                       

@article{lew2022safe,
  title={Safe active dynamics learning and control: A sequential exploration--exploitation framework},
  author={Lew, Thomas and Sharma, Apoorva and Harrison, James and Bylard, Andrew and Pavone, Marco},
  journal={IEEE Transactions on Robotics},
  volume={38},
  number={5},
  pages={2888--2907},
  year={2022},
  publisher={IEEE}
}

@article{limon2018nonlinear,
  title={Nonlinear {MPC} for tracking piece-wise constant reference signals},
  author={Limon, Daniel and Ferramosca, Antonio and Alvarado, Ignacio and Alamo, Teodoro},
  journal={IEEE Transactions on Automatic Control},
  volume={63},
  number={11},
  pages={3735--3750},
  year={2018},
  publisher={IEEE}
}

@inproceedings{harrison2018meta,
  title={Meta-learning priors for efficient online bayesian regression},
  author={Harrison, James and Sharma, Apoorva and Pavone, Marco},
  booktitle={International Workshop on the Algorithmic Foundations of Robotics},
  pages={318--337},
  year={2018},
  organization={Springer}
}

@article{prajapat2025safe,
  title={Safe guaranteed exploration for non-linear systems},
  author={Prajapat, Manish and K{\"o}hler, Johannes and Turchetta, Matteo and Krause, Andreas and Zeilinger, Melanie N},
  journal={IEEE Transactions on Automatic Control},
  year={2025},
  publisher={IEEE}
}

@article{gawlikowski2023survey,
  title={A survey of uncertainty in deep neural networks},
  author={Gawlikowski, Jakob and Tassi, Cedrique Rovile Njieutcheu and Ali, Mohsin and Lee, Jongseok and Humt, Matthias and Feng, Jianxiang and Kruspe, Anna and Triebel, Rudolph and Jung, Peter and Roscher, Ribana and others},
  journal={Artificial Intelligence Review},
  volume={56},
  number={Suppl 1},
  pages={1513--1589},
  year={2023},
  publisher={Springer}
}

@article{jospin2022hands,
  title={Hands-on Bayesian neural networks—A tutorial for deep learning users},
  author={Jospin, Laurent Valentin and Laga, Hamid and Boussaid, Farid and Buntine, Wray and Bennamoun, Mohammed},
  journal={IEEE Computational Intelligence Magazine},
  volume={17},
  number={2},
  pages={29--48},
  year={2022},
  publisher={IEEE}
}

@article{fiedler2023improved,
  title={Improved uncertainty quantification for neural networks with bayesian last layer},
  author={Fiedler, Felix and Lucia, Sergio},
  journal={IEEE Access},
  volume={11},
  pages={123149--123160},
  year={2023},
  publisher={IEEE}
}

@article{tran2019bayesian,
  title={Bayesian layers: A module for neural network uncertainty},
  author={Tran, Dustin and Dusenberry, Mike and Van Der Wilk, Mark and Hafner, Danijar},
  journal={Advances in neural information processing systems},
  volume={32},
  year={2019}
}

@inproceedings{fiedler2022model,
  title={Model predictive control with neural network system model and Bayesian last layer trust regions},
  author={Fiedler, Felix and Lucia, Sergio},
  booktitle={2022 IEEE 17th International Conference on Control \& Automation (ICCA)},
  pages={141--147},
  year={2022},
  organization={IEEE}
}

@article{pohlodek2023stochastic,
  title={Stochastic model predictive control utilizing bayesian neural networks},
  author={Pohlodek, Johannes and Alsmeier, H and Morabito, Bruno and Schlauch, Christian and Savchenko, Anton and Findeisen, Rolf},
  journal={arXiv preprint arXiv:2303.14519},
  year={2023}
}

@article{johnson2024robust,
  title={Robust nonlinear model predictive control of continuous crystallization using Bayesian last layer surrogate models},
  author={Johnson, Collin R and Fiedler, Felix and Lucia, Sergio},
  journal={IFAC-PapersOnLine},
  volume={58},
  number={14},
  pages={476--481},
  year={2024},
  publisher={Elsevier}
}

@inproceedings{jain2018learning,
  title={Learning and control using Gaussian processes},
  author={Jain, Achin and Nghiem, Truong and Morari, Manfred and Mangharam, Rahul},
  booktitle={2018 ACM/IEEE 9th international conference on cyber-physical systems (ICCPS)},
  pages={140--149},
  year={2018},
  organization={IEEE}
}

@article{mesbah2018stochastic,
  title={Stochastic model predictive control with active uncertainty learning: A survey on dual control},
  author={Mesbah, Ali},
  journal={Annual Reviews in Control},
  volume={45},
  pages={107--117},
  year={2018},
  publisher={Elsevier}
}

@article{li2024dual,
  title={Dual control of exploration and exploitation for auto-optimization control with active learning},
  author={Li, Zhongguo and Chen, Wen-Hua and Yang, Jun and Yan, Yunda},
  journal={IEEE Transactions on Automation Science and Engineering},
  volume={22},
  pages={2145--2158},
  year={2024},
  publisher={IEEE}
}

@article{garcia1989model,
	title={Model predictive control: Theory and practice—A survey},
	author={Garcia, Carlos E and Prett, David M and Morari, Manfred},
	journal={Automatica},
	volume={25},
	number={3},
	pages={335--348},
	year={1989},
	publisher={Elsevier}
}

@article{james1994certainty,
	title={On the certainty equivalence principle and the optimal control of partially observed dynamic games},
	author={James, Matthew R},
	journal={IEEE Transactions on Automatic Control},
	volume={39},
	number={11},
	pages={2321--2324},
	year={1994},
	publisher={IEEE}
}

@inproceedings{chowdhury2017kernelized,
  title={On kernelized multi-armed bandits},
  author={Chowdhury, Sayak Ray and Gopalan, Aditya},
  booktitle={International Conference on Machine Learning},
  pages={844--853},
  year={2017},
  organization={PMLR}
}

@article{bonassi2022recurrent,
  title={On recurrent neural networks for learning-based control: recent results and ideas for future developments},
  author={Bonassi, Fabio and Farina, Marcello and Xie, Jing and Scattolini, Riccardo},
  journal={Journal of Process Control},
  volume={114},
  pages={92--104},
  year={2022},
  publisher={Elsevier}
}

@article{heirung2015mpc,
  title={{MPC}-based dual control with online experiment design},
  author={Heirung, Tor Aksel N and Foss, Bjarne and Ydstie, B Erik},
  journal={Journal of Process Control},
  volume={32},
  pages={64--76},
  year={2015},
  publisher={Elsevier}
}

@article{la2023optimal,
	title={Optimal management and data-based predictive control of district heating systems: The {N}ovate {M}ilanese experimental case-study},
	author={La Bella, Alessio and Del Corno, Ada},
	journal={Control Engineering Practice},
	volume={132},
	pages={105429},
	year={2023},
	publisher={Elsevier}
}

@article{krug2021nonlinear,
	title={Nonlinear optimization of district heating networks},
	author={Krug, Richard and Mehrmann, Volker and Schmidt, Martin},
	journal={Optimization and Engineering},
	volume={22},
	number={2},
	pages={783--819},
	year={2021},
	publisher={Springer}
}

@misc{ssnet,
  author = {Bonassi, Fabio},
  title = {ssnet: a Python module for training State Space neural NETworks},
  publisher = {GitHub},
  journal = {GitHub repository},
  howpublished = {\mbox{\url{https://github.com/bonassifabio/ssnet}}},
	year={2023}
}

@article{morari1988model,
  title={Model predictive control: Theory and practice},
  author={Morari, Manfred and Garcia, Carlos E and Prett, David M},
  journal={IFAC Proceedings Volumes},
  volume={21},
  number={4},
  pages={1--12},
  year={1988},
  publisher={Elsevier}
}

@inproceedings{de2024lifelong,
  title={Lifelong learning for monitoring and adaptation of data-based dynamical models: a statistical process control approach},
  author={Boca de Giuli, Laura and La Bella, Alessio and De Nicolao, Giuseppe and Scattolini, Riccardo},
  booktitle={2024 European Control Conference (ECC)},
  pages={947--952},
  year={2024},
  organization={IEEE}
}

@book{zhang2012ensemble,
	title={Ensemble Machine Learning: Methods and Applications},
	author={Zhang, Cha and Ma, Yunqian},
	year={2012},
	publisher={Springer Science \& Business Media}
}

@article{soloperto2020augmenting,
  title={Augmenting {MPC} schemes with active learning: Intuitive tuning and guaranteed performance},
  author={Soloperto, Raffaele and K{\"o}hler, Johannes and Allg{\"o}wer, Frank},
  journal={IEEE Control Systems Letters},
  volume={4},
  number={3},
  pages={713--718},
  year={2020},
  publisher={IEEE}
}

@article{heirung2018model,
  title={Model predictive control with active learning under model uncertainty: Why, when, and how},
  author={Heirung, Tor Aksel N and Paulson, Joel A and Lee, Shinje and Mesbah, Ali},
  journal={AIChE Journal},
  volume={64},
  number={8},
  pages={3071--3081},
  year={2018},
  publisher={Wiley Online Library}
}

@article{camilleri2022active,
  title={Active learning with safety constraints},
  author={Camilleri, Romain and Wagenmaker, Andrew and Morgenstern, Jamie H and Jain, Lalit and Jamieson, Kevin G},
  journal={Advances in Neural Information Processing Systems},
  volume={35},
  pages={33201--33214},
  year={2022}
}

@article{srinivas2012information,
  title={Information-theoretic regret bounds for gaussian process optimization in the bandit setting},
  author={Srinivas, Niranjan and Krause, Andreas and Kakade, Sham M and Seeger, Matthias W},
  journal={IEEE transactions on information theory},
  volume={58},
  number={5},
  pages={3250--3265},
  year={2012},
  publisher={IEEE}
}

@ARTICLE{nigro2024control,
  author={Nigro, Lorenzo and La Bella, Alessio and Casella, Francesco and Scattolini, Riccardo},
  journal={IEEE Transactions on Automation Science and Engineering}, 
  title={Control-Oriented Modeling, Simulation, and Predictive Control of District Heating Networks}, 
  year={2025},
  volume={22},
  number={},
  pages={7064-7079}
  }

@article{prajapat2025safe2,
  title={Safe and Near-Optimal Control with Online Dynamics Learning},
  author={Prajapat, Manish and K{\"o}hler, Johannes and Zeilinger, Melanie N and Krause, Andreas},
  journal={arXiv preprint arXiv:2509.16650},
  year={2025}
}

@article{de2024physics,
  title={Physics-informed neural network modeling and predictive control of district heating systems},
  author={Boca de Giuli, Laura and La Bella, Alessio and Scattolini, Riccardo},
  journal={IEEE Transactions on Control Systems Technology},
  volume={32},
  number={4},
  pages={1182--1195},
  year={2024},
  publisher={IEEE}
}

@article{fortunato2017bayesian,
  title={Bayesian recurrent neural networks},
  author={Fortunato, Meire and Blundell, Charles and Vinyals, Oriol},
  journal={arXiv preprint arXiv:1704.02798},
  year={2017}
}

@article{cohn1993neural,
  title={Neural network exploration using optimal experiment design},
  author={Cohn, David},
  journal={Advances in neural information processing systems},
  volume={6},
  year={1993}
}

@article{ewen2023not,
  title={Not All Actions Are Created Equal: Bayesian Optimal Experimental Design for Safe and Optimal Nonlinear System Identification},
  author={Ewen, Parker and Gunjal, Gitesh and Wilson, Joey and Liu, Jinsun and Adu, Challen Enninful and Vasudevan, Ram},
  journal={arXiv preprint arXiv:2308.01829},
  year={2023}
}

@article{schneider1996exploiting,
  title={Exploiting model uncertainty estimates for safe dynamic control learning},
  author={Schneider, Jeff},
  journal={Advances in neural information processing systems},
  volume={9},
  year={1996}
}

@book{williams2006gaussian,
  title={Gaussian processes for machine learning},
  author={Williams, Christopher KI and Rasmussen, Carl Edward},
  volume={2},
  number={3},
  year={2006},
  publisher={MIT press Cambridge, MA}
}

@article{scampicchio2025gaussian,
  title={Gaussian processes for dynamics learning in model predictive control},
  author={Scampicchio, Anna and Arcari, Elena and Lahr, Amon and Zeilinger, Melanie N},
  journal={Annual Reviews in Control},
  volume={60},
  pages={101034},
  year={2025},
  publisher={Elsevier}
}

@article{medsker2001recurrent,
  title={Recurrent neural networks},
  author={Medsker, Larry R and Jain, Lakhmi and others},
  journal={Design and applications},
  volume={5},
  number={64-67},
  pages={2},
  year={2001}
}

@book{anderson1995introduction,
  title={An introduction to neural networks},
  author={Anderson, James A},
  year={1995},
  publisher={MIT press}
}

@inproceedings{vakili2021information,
  title={On information gain and regret bounds in gaussian process bandits},
  author={Vakili, Sattar and Khezeli, Kia and Picheny, Victor},
  booktitle={International Conference on Artificial Intelligence and Statistics},
  pages={82--90},
  year={2021},
  organization={PMLR}
}

@article{ravasio2024lmi,
  title={{LMI}-based design of a robust model predictive controller for a class of Recurrent Neural Networks with guaranteed properties},
  author={Ravasio, Daniele and Farina, Marcello and Ballarino, Andrea},
  journal={IEEE Control Systems Letters},
  volume={8},
  pages={1126--1131},
  year={2024},
  publisher={IEEE}
}

@article{schimperna2024robust,
  title={Robust offset-free constrained model predictive control with long short-term memory networks},
  author={Schimperna, Irene and Magni, Lalo},
  journal={IEEE Transactions on Automatic Control},
  volume={69},
  number={12},
  pages={8172--8187},
  year={2024},
  publisher={IEEE}
}

@article{bonassi2021nonlinear,
  title={Nonlinear {MPC} for offset-free tracking of systems learned by GRU neural networks},
  author={Bonassi, Fabio and da Silva, Caio Fabio Oliveira and Scattolini, Riccardo},
  journal={IFAC-PapersOnLine},
  volume={54},
  number={14},
  pages={54--59},
  year={2021},
  publisher={Elsevier}
}

@article{morato2024data,
  title={Data Science and Model Predictive Control:: A survey of recent advances on data-driven {MPC} algorithms},
  author={Morato, Marcelo M and Felix, Monica S},
  journal={Journal of Process Control},
  volume={144},
  pages={103327},
  year={2024},
  publisher={Elsevier}
}

@article{mckinnon2019learn,
  title={Learn fast, forget slow: Safe predictive learning control for systems with unknown and changing dynamics performing repetitive tasks},
  author={McKinnon, Christopher D and Schoellig, Angela P},
  journal={IEEE Robotics and Automation Letters},
  volume={4},
  number={2},
  pages={2180--2187},
  year={2019},
  publisher={IEEE}
}

@article{lorenzen2019robust,
  title={Robust {MPC} with recursive model update},
  author={Lorenzen, Matthias and Cannon, Mark and Allg{\"o}wer, Frank},
  journal={Automatica},
  volume={103},
  pages={461--471},
  year={2019},
  publisher={Elsevier}
}

@article{thangavel2018dual,
  title={Dual robust nonlinear model predictive control: A multi-stage approach},
  author={Thangavel, S and Lucia, S and Paulen, R and Engell, S},
  journal={Journal of process control},
  volume={72},
  pages={39--51},
  year={2018},
  publisher={Elsevier}
}

@article{iannelli2020structured,
  title={Structured exploration in the finite horizon linear quadratic dual control problem},
  author={Iannelli, Andrea and Khosravi, Mohammad and Smith, Roy S},
  journal={IFAC-PapersOnLine},
  volume={53},
  number={2},
  pages={959--964},
  year={2020},
  publisher={Elsevier}
}

@article{prajapat2022near,
  title={Near-Optimal Multi-Agent Learning for Safe Coverage Control},
  author={Prajapat, Manish and Turchetta, Matteo and Zeilinger, Melanie and Krause, Andreas},
  journal={Advances in Neural Information Processing Systems},
  volume={35},
  pages={14998--15012},
  year={2022}
}

@book{rawlings2017model,
  title={Model Predictive Control: Theory, Computation, and Design},
  author={Rawlings, J.B. and Mayne, D.Q. and Diehl, M.},
  isbn={9780975937730},
  url={\url{https://books.google.it/books?id=MrJctAEACAAJ}},
  year={2017},
  publisher={Nob Hill Publishing}
}

@book{boyd1994linear,
  title={Linear matrix inequalities in system and control theory},
  author={Boyd, Stephen and El Ghaoui, Laurent and Feron, Eric and Balakrishnan, Venkataramanan},
  year={1994},
  publisher={SIAM}
}

@article{liu2020gaussian,
  title={When Gaussian process meets big data: A review of scalable GPs},
  author={Liu, Haitao and Ong, Yew-Soon and Shen, Xiaobo and Cai, Jianfei},
  journal={IEEE Transactions on neural networks and learning systems},
  volume={31},
  number={11},
  pages={4405--4423},
  year={2020},
  publisher={IEEE}
}

\appendix
\section{APPENDIX}
In the following subsections, we report the proofs of Lemma~\ref{lemma:beta_k} and Theorem~\ref{th:1}. 

\subsection{Proof of Lemma~\ref{lemma:beta_k}}
\label{appendix:beta}

The proof draws inspiration from~\cite[Theorem 1]{lew2022safe}. In detail, we need to prove that, with probability at least $1-\delta$ and for all $k \in \mathbb{N}_{>0}$:
\begin{equation}
	\begin{aligned}
    & \lvert \theta^{\star \top} x -\bar{\theta}_k^{\top} x \rvert \leq \beta_{k} \Sigma_k(x).
	\end{aligned}
\end{equation}
To do so, from~\eqref{subeq:thetak} we obtain the update for $\bar{\theta}_k$:
\begin{equation}
	\begin{aligned}
    \bar{\theta}_k = \Lambda_k^{-1} (x_k \tilde{y}_k^{\star} + x_{k-1} \tilde{y}_{k-1}^{\star} + \hdots  + \Lambda_0 \bar{\theta}_0).
	\end{aligned}
\end{equation}

Now we define the auxiliary matrices storing measured state, output, and noise sequences as 
$ X_k = [x_1 \hdots x_k]^{\top} \in \mathbb{R}^{k \times n_x} $, $ \tilde{Y}_k^{\star} = [\tilde{y}_1^{\star} \hdots \tilde{y}_k^{\star}]^{\top} \in  \mathbb{R}^{k \times n_y}$, $ \Xi_k = [\eta_1 \hdots \eta_k]^{\top} \in \mathbb{R}^{k \times n_y}$, respectively. Recalling~\eqref{eq:measuredoutput}, these satisfy $\tilde{Y}_k^{\star} = X_k \theta^{\star} + \Xi_k$. We can thus write
\begin{align}
    \label{eq:thetak}
    \bar{\theta}_k = \Lambda_k^{-1} (X_k^{\top} \tilde{Y}_k^{\star} + \Lambda_0 \bar{\theta}_0) = \Lambda_k^{-1} (X_k^{\top} (X_k \theta^{\star} + \Xi_k) + \Lambda_0 \bar{\theta}_0). 
\end{align}
Now, considering that $\Lambda_k^{-1}$ is symmetric by definition of covariance matrix,~\eqref{subeq:lambdakinv} can be written as 
\begin{equation}
	\begin{aligned}
    \label{eq:lambdakinvsymm}
    \Lambda_{k}^{-1} = \Lambda_{k-1}^{-1} - \frac{\Lambda_{k-1}^{-1} x_k x_k^{\top} \Lambda_{k-1}^{-1} } {1+x_k^{\top} \Lambda_{k-1}^{-1} x_k}.
	\end{aligned}
\end{equation}
Using the Woodbury matrix identity,~\eqref{eq:lambdakinvsymm} becomes $\Lambda_{k}^{-1} = (\Lambda_{k-1} + x_k x_k^{\top})^{-1}$, which implies that 
\begin{equation}
	\begin{aligned}
    \label{eq:lambdak}
    \Lambda_{k} = \Lambda_0 + X_k^{\top} X_k.
	\end{aligned}
\end{equation}
From~\eqref{eq:thetak} we get:
\begin{align}
    \bar{\theta}_k & = \Lambda_k^{-1} X_k^{\top} \Xi_k + \Lambda_k^{-1} X_k^{\top} X_k \theta^{\star} + \Lambda_k^{-1} \Lambda_0 \bar{\theta}_0 \nonumber \\
    & \stackrel{(i)}{=} \Lambda_k^{-1} X_k^{\top} \Xi_k + \Lambda_k^{-1}(X_k^{\top} X_k + \Lambda_0)\theta^{\star} + \Lambda_k^{-1} \Lambda_0(\bar{\theta}_0-\theta^{\star}) \nonumber \\
    & \stackrel{(ii)}{=} \Lambda_k^{-1} X_k^{\top} \Xi_k + \theta^{\star} - \Lambda_k^{-1} \Lambda_0 (\theta^{\star}-\bar{\theta}_0),
\end{align}
where step $(i)$ follows by adding and subtracting $\Lambda_k^{-1} \Lambda_0 \theta^{\star}$ and $(ii)$ by substituting~\eqref{eq:lambdak}. Now, for any $a \in \mathbb{R}^{n_x}$, we can write
\begin{align}
    \label{eq:a}
    a^{\top}(\bar{\theta}_k-\theta^{\star}) & = a^{\top}(\Lambda_k^{-1} X_k^{\top} \Xi_k) - a^{\top}(\Lambda_k^{-1} \Lambda_0 (\theta^{\star}-\bar{\theta}_0)), \nonumber \\
    \lvert a^{\top}(\bar{\theta}_k-\theta^{\star}) \rvert & \stackrel{(i)}{\leq} \lVert a \rVert_{\Lambda_k^{-1}} (\lVert X_k^{\top} \Xi_k \rVert_{\Lambda_k^{-1}} + \lVert \Lambda_0 (\theta^{\star}-\bar{\theta}_0) \rVert_{\Lambda_k^{-1}})  \stackrel{(ii)}{\leq} \lVert a \rVert_{\Lambda_k^{-1}} (\lVert X_k^{\top} \Xi_k \rVert_{\Lambda_k^{-1}} + \lVert \theta^{\star}-\bar{\theta}_0 \rVert_{\Lambda_0} ),
\end{align}
where $(i)$ comes from the Cauchy-Schwarz inequality, and $(ii)$ follows from 
\begin{align}
    \lVert \Lambda_0 (\theta^{\star}-\bar{\theta}_0) \rVert_{\Lambda_k^{-1}}^2  \leq \lambda_{\mathrm{max}}(\Lambda_k^{-1},\Lambda_0^{-1})\lVert  \theta^{\star}-\bar{\theta}_0 \rVert_{\Lambda_0}^2  \leq \lVert  \theta^{\star}-\bar{\theta}_0 \rVert_{\Lambda_0}^2,
\end{align}
with $\lambda_{\mathrm{max}}$ denoting the maximum generalised eigenvalue of $\Lambda_k^{-1}$ and $\Lambda_0^{-1}$, which is not larger than 1 since $\Lambda_0^{-1}\succeq\Lambda_k^{-1}$, as evident from~\eqref{subeq:lambdakinv},~\citep{boyd1994linear}.
Now, using~\citep[Lemma A.1]{lew2022safe}, with probability at least $1-\delta$ it holds that
\begin{equation}
	\begin{aligned}
    \lVert X_k^{\top} \Xi_k \rVert_{\Lambda_k^{-1}}^2 \leq 2 \sigma^2 \log \left ( \frac{1}{\delta} \frac{\det (\Lambda_k)^{1/2}}{\det (\Lambda_0)^{1/2}} \right ).
	\end{aligned}
\end{equation}
Then, considering Assumption~\ref{ass:prior}, we have $\lVert \theta^{\star} - \bar{\theta}_0 \rVert_{\Lambda_0}^2 \leq C$. Therefore, considering~\eqref{eq:a}, with probability at least $1-\delta$, we get
\begin{align}
    \label{eq:aT}
    \lvert a^{\top}( \bar{\theta}_k-\theta^{\star}) \rvert \leq  \lVert a \rVert_{\Lambda_k^{-1}} \sigma \left (  \sqrt{2 \log \left( \frac{1}{\delta} \frac{\det(\Lambda_k)^{1/2}}{\det(\Lambda_0)^{1/2}} \right)} + \sqrt{\frac{C}{\sigma^2}} \right) = \lVert a \rVert_{\Lambda_k^{-1}} \sigma \beta_{k}.
\end{align}
Setting $a=x$, we get from~\eqref{eq:aT}:
\begin{equation}
    \begin{aligned}
    & \lvert x^{\top} \bar{\theta}_k- x^{\top} \theta^{\star} \rvert \leq \lVert x \rVert_{\Lambda_k^{-1}} \cdot \sigma \cdot \beta_{k}. 
    \end{aligned}
\end{equation}
From~\eqref{subeq:Sigmak2}, we have that $\Sigma_k(x)= \sigma \sqrt{x^{\top} \Lambda_k^{-1} x} $, and thus the claim follows. $\hfill \square$

\subsection{Proof of Theorem~\ref{th:1}}
\label{appendix:prooftheo1}
The proof builds on preliminary results discussed in~\citep{prajapat2025safe2}.

(1) \textit{Recursive feasibility}. Consider Algorithm~\ref{algo}. At $k=1$, the pessimistic~\eqref{eq:pessimisticMPC} and the optimistic~\eqref{eq:optimisticMPC} problems are feasible since $x_1 \in \mathbb{X}_1$ and due to the control invariance property of Assumption~\ref{ass:invset}. The exploration problem~\eqref{eq:safeexploMPC} is feasible for the same reason, and also because of the slack variables in~\eqref{subeq:explo_w}-\eqref{subeq:explo_nu}. For $k > 1$, we note from~\eqref{eq:bounds}-\eqref{eq:Xkp} that $\mathcal{X}_k^{\mathrm{p}} \subseteq \mathcal{X}_{k+1}^{\mathrm{p}} $: thanks to this and to Assumption~\ref{ass:invset}, common arguments in MPC theory can be applied~\citep{morari1988model} and recursive feasibility is proved for the three optimisation problems. $\hfill \square$ 

(2) \textit{Safety}. Thanks to~\eqref{subeq:explo_u} and~\eqref{subeq:upess}, the control actions $u_{0:h^{\star}-1\mid k}^{\mathrm{e}}$ from problem~\eqref{eq:safeexploMPC} and $u_{0 \mid k}^{\mathrm{p}}$ from~\eqref{eq:pessimisticMPC}, respectively, ensure that $\bar{\theta}_k^{\top} x \in \mathcal{Y} = [y_{\mathrm{min}}, y_{\mathrm{max}}]$ for all $x \in \mathcal{X}_k^{\mathrm{p}}$. By Corollary~\ref{corollary:beta}, it holds that $\theta^{\star \top} x \in \mathcal{Y}$ as well with probability at least $1-\delta$. $\hfill \square$ 

(3) \textit{Finite termination of exploration}. First, we prove that there exists a finite $n^{\star} \in \mathbb{N}$ such that $w_{k_{n^{\star}}}(x_{h \mid k_{n^{\star}}}) < \epsilon$, for each $x_{0:H \mid k_{n^{\star}}}$, where $x_{0 \mid k_{n^{\star}}}=x_{k_{n^{\star}}}$, $x_{h+1 \mid k_{n^{\star}}} = \phi(x_{h \mid k_{n^{\star}}}, u_{h \mid k_{n^{\star}}})$, $u_{0:H-1 \mid k_{n^{\star}}} \in \mathcal{U}^{\mathrm{p}}_{k_{n^{\star}}}(x_{k_{n^{\star}}})$. Note that, in view of Assumption~\ref{ass:exactpenalty}, if an $\epsilon$-informative state can be reached along the prediction horizon (i.e., if~\eqref{subeq:explo_w} can be satisfied with a zero slack variable at some instants), the exploration problem \eqref{eq:safeexploMPC} will not assign a non-zero value to the slack variable. 

Therefore, we notice that the confidence width $w_{k_{i-1}}^2(x_{k_{i}})$ $=\beta_{k_{i-1}}^2 \Sigma_{k_{i-1}}^2(x_{k_{i}})$ can be bounded as:
\begin{align}
    w_{k_{i-1}}^2(x_{k_{i}}) \leq \sum_{k=k_{i-1}}^{k_{i}} \beta_{k_{i-1}}^2 \Sigma_{k_{i-1}}^2(x_{k}) = \beta_{k_{i-1}}^2 \sum_{k=k_{i-1}}^{k_{i}} x_k^{\top} \sigma^2 \Lambda_{k_{i-1}}^{-1} x_k  \stackrel{(i)}{\leq} \beta_{k_{i-1}}^2 C_1 \frac{1}{2} \sum_{k=k_{i-1}}^{k_{i}} \log\Big(  1+ x_k^{\top} \Lambda_{k_{i-1}}^{-1} x_k \Big),
\end{align}
where $(i)$ is obtained using similar steps to~\citep[Lemma 7]{prajapat2022near}, with $C_1 = \frac{2 H}{\log(1+H \sigma^{-2})}$. Moreover, considering that we collect $\epsilon$-informative states at each $k_i$, it holds that
\begin{align}
    \sum_{i=1}^{n} w_{k_{i-1}}^2(x_{k_{i}}) \geq n \epsilon^2,
\end{align}
and thus
\begin{align}
\label{eq:nstar}
    n \epsilon^2 \leq \sum_{i=1}^{n} \beta_{k_{i-1}}^2 C_1 \frac{1}{2} \sum_{k=k_{i-1}}^{k_{i}} \log\Big(  1+ x_k^{\top} \Lambda_{k_{i-1}}^{-1} x_k \Big)  \stackrel{(i)}{\leq} \beta_{k_n}^2 C_1 \sum_{i=1}^{n} \frac{1}{2} \sum_{k=k_{i-1}}^{k_{i}} \log\Big(  1+ x_k^{\top} \Lambda_{k_{i-1}}^{-1} x_k \Big) \stackrel{(ii)}{\leq} \beta_{k_n}^2 C_1 \gamma_{k_n},
\end{align}
where $(i)$ follows since $\beta_{k_n}$~\eqref{eq:beta} is non-decreasing with $k_n$, and $\gamma_{k_n}$ in $(ii)$ can be defined as the upper bound of the information capacity~\citep[Lemma 5.4]{srinivas2012information}. Noting that $\beta_{k_n}^2 \gamma_{k_n}$ grows sublinearly with $k_n$ in the considered case of finite-dimensional parameters $\theta$ and linear kernels~\citep{srinivas2012information,vakili2021information}, it follows that there exists a largest integer $n^{\star}$ satisfying~\eqref{eq:nstar} which, consequently, implies that $w_{k_{n^{\star}}}(x_{h \mid k_{n^{\star}}}) < \epsilon$, for each $x_{0:H \mid k_{n^{\star}}}$, where $x_{0 \mid k_{n^{\star}}}=x_{k_{n^{\star}}}$, $x_{h+1 \mid k_{n^{\star}}} = \phi(x_{h \mid k_{n^{\star}}}, u_{h \mid k_{n^{\star}}})$, $u_{0:H-1 \mid k_{n^{\star}}} \in \mathcal{U}^{\mathrm{p}}_{k_{n^{\star}}}(x_{k_{n^{\star}}})$.

Second, we prove that this condition implies that
\begin{align}
    \mathcal{U}_{k_{n^{\star}}}^{\mathrm{o},2 \epsilon}(x_{k_{n^{\star}}}) \subseteq \mathcal{U}_{k_{n^{\star}}}^{\mathrm{p}}(x_{k_{n^{\star}}}).
\end{align}
We prove this by contradiction, assuming that there exists $u^{\dagger}_{0:H-1 \mid k_{n^{\star}}} \in \mathcal{U}_{k_{n^{\star}}}^{\mathrm{o},2 \epsilon}(x_{k_{n^{\star}}}) \backslash \mathcal{U}_{k_{n^{\star}}}^{\mathrm{p}}(x_{k_{n^{\star}}}) $ arbitrarily close to the boundary of $\mathcal{U}_{k_{n^{\star}}}^{\mathrm{p}}(x_{k_{n^{\star}}})$, that is, there exists $u^{\mathrm{p}}_{0:H-1 \mid k_{n^{\star}}} \in \mathcal{U}_{k_{n^{\star}}}^{\mathrm{p}}(x_{k_{n^{\star}}})$ such that
\begin{align}
\label{eq:epsilonu}
    \lVert u^{\dagger}_{0:H-1 \mid k_{n^{\star}}}- u^{\mathrm{p}}_{0:H-1 \mid k_{n^{\star}}} \rVert \leq \epsilon_u, 
\end{align}
with $\epsilon_u \in \mathbb{R}_{>0}$ arbitrarily small. This implies that $w_{k_{n^{\star}}}(x_{h \mid k_{n^{\star}}}) < \epsilon$, with $x_{0 \mid k_{n^{\star}}}=x_{k_{n^{\star}}}$, $x_{h+1 \mid k_{n^{\star}}} = \phi(x_{h \mid k_{n^{\star}}}, u^{\dagger}_{h \mid k_{n^{\star}}})$, $h \in [0,H-1]$. Since $u^{\dagger}_{0:H-1 \mid k_{n^{\star}}} \in \mathcal{U}_{k_{n^{\star}}}^{\mathrm{o},2 \epsilon}(x_{k_{n^{\star}}})$, then there exists $\theta^{\mathrm{o}} \in \Theta_{k_{n^{\star}}}$ such that, given~\eqref{eq:Xo}-\eqref{eq:Uo},
\begin{equation}
	\label{eq:theta0bound1}
	\left\{
	\begin{aligned}
		\theta^{\mathrm{o}^{\top}} x_{h \mid k_{n^{\star}}} \leq y_{\mathrm{max}} - 2 \epsilon \\
        \theta^{\mathrm{o}^{\top}} x_{h \mid k_{n^{\star}}} \geq y_{\mathrm{min}} + 2 \epsilon
	\end{aligned},
	\right.
\end{equation}
with $x_{0 \mid k_{n^{\star}}}=x_{k_{n^{\star}}}$ and $x_{h+1 \mid k_{n^{\star}}}=\phi(x_{h \mid k_{n^{\star}}},u_{h \mid k_{n^{\star}}}^{\dagger}), \, h \in [0,H-1]$. Moreover, we have that for all $\theta \in \Theta_{k_{n^{\star}}}$, since also $\theta^{\mathrm{o}} \in \Theta_{k_{n^{\star}}}$,
\begin{align}
    \lvert \theta^{\top} x_{h \mid k_{n^{\star}}} - \theta^{\mathrm{o}^{\top}} x_{h \mid k_{n^{\star}}} \rvert \leq 2 w_{k_{n^{\star}}}(x_{h \mid k_{n^{\star}}}) < 2 \epsilon,
\end{align}
with $x_{0 \mid k_{n^{\star}}}=x_{k_{n^{\star}}}$ and $x_{h+1 \mid k_{n^{\star}}}=\phi(x_{h \mid k_{n^{\star}}},u_{h \mid k_{n^{\star}}}^{\dagger}), \, h \in [0,H-1]$. This implies that 
\begin{equation}
	\label{eq:theta0bound2}
	\left\{
	\begin{aligned}
		\theta^{\mathrm{o}^{\top}} x_{h \mid k_{n^{\star}}} < \theta^{\top} x_{h \mid k_{n^{\star}}} + 2 \epsilon \\
        \theta^{\mathrm{o}^{\top}} x_{h \mid k_{n^{\star}}} > \theta^{\top} x_{h \mid k_{n^{\star}}} - 2 \epsilon
	\end{aligned}.
	\right.
\end{equation}
Combining~\eqref{eq:theta0bound1} and~\eqref{eq:theta0bound2}, we get 
\begin{equation}
	\label{eq:theta0bound3}
	\left\{
	\begin{aligned}
		\theta^{\top} x_{h \mid k_{n^{\star}}} - 2 \epsilon < \theta^{\mathrm{o}^{\top}} x_{h \mid k_{n^{\star}}} \leq y_{\mathrm{max}} - 2 \epsilon \\
        y_{\mathrm{min}} + 2 \epsilon \leq \theta^{\mathrm{o}^{\top}} x_{h \mid k_{n^{\star}}} < \theta^{\top} x_{h \mid k_{n^{\star}}} + 2 \epsilon
	\end{aligned},
	\right.
\end{equation}
which implies that for all $\theta \in \Theta_{k_{n^{\star}}}$, it holds that $y_{\mathrm{min}} < \theta^{\top} x_{h \mid k_{n^{\star}}} < y_{\mathrm{max}} $, and therefore $u^{\dagger}_{0:H-1 \mid k_{n^{\star}}} \in \mathcal{U}_{k_{n^{\star}}}^{\mathrm{p}}(x_{k_{n^{\star}}})$, leading to a contradiction. Consequently, $\mathcal{U}_{k_{n^{\star}}}^{\mathrm{o},2 \epsilon}(x_{k_{n^{\star}}}) \allowbreak\backslash\allowbreak \mathcal{U}_{k_{n^{\star}}}^{\mathrm{p}}(x_{k_{n^{\star}}}) = \emptyset$. \\
To conclude the proof, we show that the condition $\mathcal{U}_{k_{n^{\star}}}^{\mathrm{o},2 \epsilon}(x_{k_{n^{\star}}}) \subseteq \mathcal{U}_{k_{n^{\star}}}^{\mathrm{p}}(x_{k_{n^{\star}}})$ implies the switching condition $J_{k_{n^{\star}}}^{\mathrm{p}} - J_{k_{n^{\star}}}^{\mathrm{o}} \leq 2 L \epsilon H$. To this end, let us consider the optimal optimistic control sequence $u^{\mathrm{o}}_{0:H-1 \mid k_{n^{\star}}} \in \mathcal{U}_{k_{n^{\star}}}^{\mathrm{o},2 \epsilon}(x_{k_{n^{\star}}})$. We can write, for all $\theta^{\mathrm{o}} \in \Theta_{k_{n^{\star}}}$,
\begin{align}
    J_{k_{n^{\star}}}(x_{k_{n^{\star}}}, \bar{\theta}_{k_{n^{\star}}}, u_{0:H-1 \mid k_{n^{\star}}}^{\mathrm{o}})-J_{k_{n^{\star}}}&(x_{k_{n^{\star}}}, \theta^{\mathrm{o}}, u_{0:H-1 \mid k_{n^{\star}}}^{\mathrm{o}}) 
    = \sum\limits_{h=0}^{H-1} \Big ( \mathcal{L}_{k_{n^{\star}}+h}(\bar{\theta}_{k_{n^{\star}}}^{\top} x_{h \mid k_{n^{\star}}}) - \mathcal{L}_{k_{n^{\star}}+h}(\theta^{\mathrm{o}^{\top}} x_{h \mid k_{n^{\star}}})\Big )  \nonumber \\
    & \stackrel{(i)}{\leq} L \sum\limits_{h=0}^{H-1} \lVert \bar{\theta}_{k_{n^{\star}}}^{\top} x_{h \mid k_{n^{\star}}} - \theta^{\mathrm{o}^{\top}} x_{h \mid k_{n^{\star}}} \rVert  \stackrel{(ii)}{\leq} L \sum\limits_{h=0}^{H-1} w_{k_{n^{\star}}}(x_{h \mid k_{n^{\star}}}) \leq L \epsilon H, \label{eq:diffJ}  
\end{align}
where $(i)$ follows from Assumption~\ref{ass:lips} and $(ii)$ from the definition of $w_k(x)=\beta_{k} \Sigma_k(x)$. Consider the optimal pessimistic control sequence $u_{0:H-1 \mid k_{n^{\star}}}^{\mathrm{p}} \in \mathcal{U}_{k_{n^{\star}}}^{\mathrm{p}}(x_{k_{n^{\star}}})$: from~\eqref{eq:diffJ} and $\mathcal{U}_{k_{n^{\star}}}^{\mathrm{o},2 \epsilon}(x_{k_{n^{\star}}}) \subseteq \mathcal{U}_{k_{n^{\star}}}^{\mathrm{p}}(x_{k_{n^{\star}}})$, it follows that
\begin{align}
    J_{k_{n^{\star}}}(x_{k_{n^{\star}}}, \bar{\theta}_{k_{n^{\star}}},  u_{0:H-1 \mid k_{n^{\star}}}^{\mathrm{p}}) \leq  J_{k_{n^{\star}}}(x_{k_{n^{\star}}}, \bar{\theta}_{k_{n^{\star}}}, u_{0:H-1 \mid k_{n^{\star}}}^{\mathrm{o}})  \leq \underbrace{J_{k_{n^{\star}}}(x_{k_{n^{\star}}}, \theta^{\mathrm{o}}, u_{0:H-1 \mid k_{n^{\star}}}^{\mathrm{o}})}_{J_{k_{n^{\star}}}^{\mathrm{o}}} + L \epsilon H.
\end{align}
Considering the cost function of problem~\eqref{eq:pessimisticMPC},
\begin{align}
   J_{k_{n^{\star}}}^{\mathrm{p}} =  J_{k_{n^{\star}}}(x_{k_{n^{\star}}}, \bar{\theta}_{k_{n^{\star}}}, u_{0:H-1\mid k_{n^{\star}}}^{\mathrm{p}}) + L \sum\limits_{h=0}^{H-1}w_{k_{n^{\star}}}(x_{h \mid k_{n^{\star}}}),
\end{align}
it follows that
\begin{align}
    J_{k_{n^{\star}}}^{\mathrm{p}} - J_{k_{n^{\star}}}^{\mathrm{o}} \leq L \epsilon H + L \sum\limits_{h=0}^{H-1}w_{k_{n^{\star}}}(x_{h \mid k_{n^{\star}}}) \leq 2 L \epsilon H, 
\end{align}
which is the switching condition. To sum up, we have proven that there exists $n^{\star}$ such that $w_{k_{n^{\star}}}(x_{h \mid k_{n^{\star}}}) < \epsilon$, which implies $\mathcal{U}_{k_{n^{\star}}}^{\mathrm{o},2 \epsilon}(x_{k_{n^{\star}}}) \subseteq \mathcal{U}_{k_{n^{\star}}}^{\mathrm{p}}(x_{k_{n^{\star}}})$, which in return implies that the switching condition $J_{k_{n^{\star}}}^{\mathrm{p}} - J_{k_{n^{\star}}}^{\mathrm{o}} \leq \xi=2 L \epsilon H$ holds, indicating that Algorithm~\ref{algo} terminates each exploration phase within at most $n^{\star}$ iterations. 
It is now evident that the $2\epsilon$-approximation of the optimistic set in \eqref{eq:Uo}, the second term in the cost function of problem \eqref{eq:pessimisticMPC}, and the switching threshold $\xi$ are carefully chosen to ensure that Algorithm \ref{algo} terminates exploration in a finite number of iterations. 
$\hfill \square$ 

(4) \textit{Close-to-optimal performance}. Consider the optimal pessimistic control sequence $u_{0:H-1 \mid k}^{\mathrm{p}} \in \mathcal{U}_{k}^{\mathrm{p}}(x_k)$ and the difference 
\begin{align}
    J_k( x_k, \theta^{\star}, u_{0:H-1 \mid k}^{\mathrm{p}})-J_k(x_k, \bar{\theta}_k, u_{0:H-1 \mid k}^{\mathrm{p}})  & = \sum\limits_{h=0}^{H-1} \mathcal{L}_{k+h}(\theta^{\star \top} x_{h \mid k}) - \sum\limits_{h=0}^{H-1} \mathcal{L}_{k+h}(\bar{\theta}_k^{\top} x_{h \mid k}) \nonumber \\
     & \stackrel{(i)}{\leq} \sum\limits_{h=0}^{H-1} L \lVert \theta^{\star \top} x_{h \mid k} - \bar{\theta}_k^{\top} x_{h \mid k} \rVert  \stackrel{(ii)}{\leq} L \sum\limits_{h=0}^{H-1} w_k(x_{h \mid k}), \label{eq:Jsmaller} 
\end{align}
where $x_{h+1 \mid k} = \phi(x_{h \mid k}, u_{h \mid k})$ with $x_{0 \mid k}=x_k$, step~$(i)$ follows from Assumption~\ref{ass:lips}, and step~$(ii)$ from Lemma~\ref{lemma:beta_k}. Defining $ J_k^{\mathrm{p}} = J_k(x_k, \bar{\theta}_k, u_{0:H-1 \mid k}^{\mathrm{p}}) + L \sum\limits_{h=0}^{H-1} w_k(x_{h \mid k}) $,
the switching condition $J^{\mathrm{p}}_k-J^{\mathrm{o}}_k \leq \xi$ can be rewritten as 
\begin{align}
    J_k(x_k, \bar{\theta}_k, u_{0:H-1 \mid k}^{\mathrm{p}}) + L \sum\limits_{h=0}^{H-1} w_k(x_{h \mid k}) \leq J_k^{\mathrm{o}} + \xi.
\end{align}
Thus, inequality~\eqref{eq:Jsmaller} can be rewritten as
\begin{align}
\label{eq:JJth}
    J_k(x_k, \theta^{\star}, u_{0:H-1 \mid k}^{\mathrm{p}}) \leq J^{\mathrm{o}}_k + \xi \stackrel{(i)}{\leq} \min_{u_{0:H-1 \mid k}\in \mathcal{U}^{\star,2 \epsilon}_k(x_k)}  J_k(x_k, \theta^{\star}, u_{0:H-1 \mid k}) + \xi,
\end{align}
where $(i)$ holds since the real set is always a subset of the optimistic set, i.e., $\mathcal{U}_k^{\star,2 \epsilon}(x_k) \subseteq \mathcal{U}_k^{\mathrm{o},2 \epsilon}(x_k)$. Ultimately, inequality~\eqref{eq:JJth} implies satisfying~\eqref{eq:closetoopt}. $\hfill \square$ 

\end{document}